\documentclass{JHEP3}

\usepackage{amsfonts,amssymb,amsthm,amsmath}

\def\cB{\mathcal{B}}

\def\cE{\mathcal{E}}

\def\cL{\mathcal{L}}
\def\cM{\mathcal{M}}
\def\cN{\mathcal{N}}

\def\cQ{\mathcal{Q}}

\def\cT{\mathcal{T}}

\def\cV{\mathcal{V}}

\def\Q#1#2{\frac{\partial #1}{\partial #2}}
\def\d{\partial}

\def\varQ#1#2{\frac{\delta #1}{\delta #2}}

\def\eps{\epsilon}
\def\vareps{\varepsilon}

\def\dH{\text{d}}

\def\half{\frac{1}{2}}

\title{Setting the boundary free in AdS/CFT}

\author{Geoffrey Comp\`ere and Donald Marolf \\
Department of Physics, University of California at Santa Barbara,
Santa Barbara, CA 93106, USA \\ E-mail:
 \it{gcompere@physics.ucsb.edu, marolf@physics.ucsb.edu }  }

\abstract{We describe a new class of boundary conditions for
AdS$_{d+1}$ under which the boundary metric becomes a dynamical
field.  The key technical point is to show that contributions from
boundary counter-terms in the bulk gravitational action render such
fluctuations normalizable. In the context of AdS/CFT, the
analogue of Neumann boundary conditions for AdS promotes the CFT
metric to a dynamical field but adds no explicit gravitational
dynamics; the gravitational dynamics is just that induced by the
conformal fields. Other AdS boundary conditions couple the CFT to a
gravity theory of choice. We use this correspondence to briefly
explore the coupled CFT + gravity theories and, in particular, for
$d=3$ we show that coupling topologically massive gravity to a large
$N$ CFT preserves the perturbative stability of the theory with
negative (3-dimensional) Newton's constant.}


\begin{document}

\section{Introduction}

The AdS/CFT correspondence was originally motivated by studying the
low energy limit of certain brane configurations in string- and
M-theory \cite{Maldacena:1998re}.  The immediate result of this
argument is a set of dualities between the specific conformal field
theories (CFTs) obtained in this limit and string/M- theory with
certain asymptotically AdS$_{d+1} \times K$ boundary conditions.
However, it quickly became clear that various deformations of the
CFT are dual to deformations of the AdS boundary conditions
\cite{Gubser:1998bc,Witten:1998qj}. This is now understood to be a
rather general principle which allows one to add new multi-trace
interactions to the CFT
\cite{Witten:Multitrace,SSandBerkooz,SeverShomer}, and even to
couple the CFT to additional dynamical fields
\cite{Witten:2003ya,Rocha}.

We shall be interested here in a deformation of AdS/CFT which
promotes the metric of the CFT to a dynamical field, coupling the
CFT to dynamical gravity.  For $d=3$ (AdS${}_4$), the possibility of
such a string/gravity duality was suggested in
\cite{Leigh:2003ez,Leigh:2007wf} (motivated by the analogy between
electro-magnetic duality for linear tensor fields and the Maxwell
field discussion of \cite{Witten:2003ya}) and in
\cite{Marolf:2006nd} (based on the observation of
\cite{Ishibashi:2004wx} that the linearized gravitational dynamics
admits a variety of self-adjoint extensions). We will argue below
that such a duality in fact holds at the non-linear level for all
dimensions $d$.

We begin by comparing the AdS and CFT path integrals along the lines
discussed in \cite{Witten:2003ya,Yee} for U(1) Maxwell fields
and in \cite{Kiritsis} in the context of cut-off CFTs. It is
convenient to use Euclidean language, but this choice of signature
plays no fundamental role in the argument. Our starting point is the
CFT partition function $Z_{CFT}[g_{(0)}]$, where $g_{(0)ij}$ is a
background metric. Promoting $g_{(0)ij}$ to a dynamical field means
that we now define the ``induced gravity'' \cite{Sakharov,Adler}
partition function by integrating $Z_{CFT}[g_{(0)}]$ over all
metrics; i.e.,

 \begin{equation}
 \label{IG}
 Z_{induced} := \int {\cal D}g_0 \  Z_{CFT}[g_{(0)}].
 \end{equation}
Note that $\ln Z^{-1}_{CFT}[g_0]$ defines the effective action for
$g_{(0)ij}$ associated with integrating out the fields of the
original CFT.  This effective action will generally include kinetic
terms for $g_{(0)ij}$, and in fact will be non-local.  Thus,
although (\ref{IG}) contains no explicit kinetic term for
$g_{(0)ij}$, both the propagator and the commutation relations for
$g_{(0)ij}$ will be non-trivial.  The metric has indeed become a
dynamical field.

Now consider performing the same operation on the bulk theory. We
impose asymptotically AdS${}_{d+1}$ boundary conditions\footnote{We
shall treat the correspondence as a duality between a theory on
AdS${}_{d+1}$ and a $d$-dimensional CFT.  We assume that any extra
compact factor $K$ in the spacetime (e.g., $S^5$, or $S^3 \times
T^4$) has been dealt with by Kaluza-Klein reduction; i.e. by
expressing the 10- or 11-dimensional fields as sums of harmonics on
$K$.  We refer only to the resulting $(d+1)$-dimensional metric in
(\ref{eq:FG1}).} which require the metric to take the form
 \begin{eqnarray}
 ds^2 &=& \frac{l^2}{r^2}dr^2 + \gamma_{ij} dx^i dx^j, \  \ \
 {\rm with} \ \label{eq:FG1}
 \cr  \gamma_{ij} &=& r^{-2} g_{(0)ij} + O(1).
 \end{eqnarray}
On-shell, the Einstein equations require
\begin{equation} \label{eq:FG2}
\gamma_{ij} = r^{-2} g_{(0)ij} +  g_{(2)ij}+\dots
 +r^{d-2} g_{(d)ij}+r^{d-2} \log r^2 \tilde g_{(d)ij}+ O(r^{d+1}), \
 \ \ \ \ \ \
 \end{equation}
 where the logarithmic term appears only for even $d$.
For the current suggestive argument, we model the bulk partition
function as a path integral over bulk fields so that the usual
AdS/CFT correspondence may be written
 \begin{equation}
 \label{AdSCFT}
 Z_{CFT}[g_{(0)ij}] = \int ({\cal D}g)_{g_{(0)}} e^{-S_{Dir}} =:
 Z_{Dir}[g_{(0)ij}].
 \end{equation}
Here $({\cal D}g)_{g_{(0)}}$ is the measure which integrates over
all metrics $g_{ij}$ corresponding via (\ref{eq:FG1}),
(\ref{eq:FG2}) to the given boundary metric $g_{(0)ij}$.  The
``Dirichlet'' action $S_{Dir} = S_{EH} + S_{GH} + S_{ct}$ contains
the usual bulk Einstein-Hilbert and cosmological constant terms
($S_{EH}$) as well as the Gibbons-Hawking boundary term ($S_{GH}$)
and the counter-terms
\cite{Henningson:1998gx,Balasubramanian:1999re,deHaro:2000xn,Papadimitriou:2005ii}
($S_{ct}$) at infinity required to make the action a well-defined
variational principle for each fixed $g_{(0)ij}$. In (\ref{AdSCFT})
we have suppressed the path integral over all non-metric bulk fields
as they play no special role in our discussion.

The induced-gravity boundary partition function (\ref{IG}) may now
be written in the form
 \begin{equation}
 \label{Neu}
 Z_{induced} =  \int {\cal D}g_0  Z_{Dir}[g_{(0)}] = \int ({\cal
 D}g)_{g_{(0)}} e^{-S_{Dir}} =: Z_{Neu},
 \end{equation}
where ${\cal D} g  =  {\cal D} g_{(0)} ({\cal D}g)_{g_{(0)}}$
integrates over {\em all} metrics of the form $(\ref{eq:FG1})$
without restriction.   In (\ref{Neu}) we have introduced a new bulk
partition function $Z_{Neu}$ which may be interpreted by passing to
the semi-classical limit. Since $S_{Dir}$ is the full bulk action
(and in particular includes all counter-terms at infinity), its
variation satisfies
 \begin{equation}
 \label{varyDir}
 \delta S_{Dir} = EOM + \frac{1}{2} \int_{\partial M} d^d x \sqrt{-g_{(0)}}
 T^{ij} \delta g_{(0)ij},
 \end{equation}
where (\ref{varyDir}) defines the boundary stress tensor $T^{ij}$
and $EOM$ represents a collection of terms that vanish when the
usual bulk equations of motion are imposed. Because $Z_{Neu}$
includes an integral over $g_{(0)ij}$, we see that stationarity of
$S_{Dir}$ in the semi-classical limit imposes the {\em Neumann}
boundary condition $T^{ij} =0$. For this reason we refer to
$Z_{Neu}$ as the Neumann partition function.  We will also find it
convenient to introduce the redundant notation $S_{Neu} = S_{EH} +
S_{GH}+ S_{ct}$ to describe the Dirichlet action when it is intended
to be used with the Neumann boundary condition $T^{ij}=0$.

The path-integral argument above  suggests that the $T^{ij} =0$
boundary condition defines a bulk AdS$_{d+1}$ theory dual to the
particular $d$-dimensional gravity theory induced from the CFT.
However, one might be concerned that fluctuations of the bulk metric
would fail to be normalizable unless they satisfy a Dirichlet
boundary condition in which $g_{(0)ij}$ is fixed (see
\cite{Ashtekar:1984,Henneaux:1985tv,Henneaux:1985ey,Brown:1986nw}
for classic references). This seems to be the conclusion reached by
\cite{BKL,BKLT,Balasubramanian:2000pq}, and is indeed the case if
one uses the usual norm associated with the Einstein-Hilbert action.
For many readers, this will be most easily seen by recalling the
results of \cite{RS1,RS2} which used an orbifold to cut-off an
AdS${}_5$ spacetime, removing the region near the boundary. In the
limit that this regulator was removed, \cite{RS1,RS2} found a
divergent norm for all modes which describe fluctuations in the
metric on the orbifold. Such results may appear to suggest that,
despite the path-integral argument above, our Neumann boundary
condition fails to define a consistent theory with dynamical
boundary metric\footnote{On the other hand, the fact that bulk AdS
methods allow one to calculate a finite effective action for
$g_{(0)}$
\cite{Liu:1998bu,Skenderis:1999nb,Hawking:2000bb,Papadimitriou:2007sj}
may be taken as evidence in favor of normalizeability. There is also
the result of \cite{Ishibashi:2004wx} noted earlier that the desired
fluctuations are normalizable for $d=3$, though it is not
immediately clear how the inner product used in
\cite{Ishibashi:2004wx} relates to the symplectic structure of AdS
gravity.}.

Our purpose here is to resolve this tension by showing for $d=2,3,4$
that modes in which $g_{(0)}$ fluctuates {\em are} in fact
normalizable with respect to the symplectic structure defined by the
full action $S_{Neu} = S_{Dir}$.  We expect this to be true for all
$d$. The key point is to take into account contributions to the norm
coming from the counterterm action $S_{ct}$, which do not seem to
have been previously considered.  Together with the boundary
condition $T_{ij}=0$, the counterterm contributions are also
essential in ensuring that the symplectic structure is conserved.

With the above understanding, the gravitational field has a status
similar to that of tachyonic scalars close to the
Breitenlohner-Freedman bound ($m_{BF}^2$)
\cite{Breitenlohner:1982bm,Breitenlohner:1982jf} and the choice of
Dirichlet vs. Neumann boundary conditions for the metric corresponds
to the two natural boundary conditions for such scalars described in
\cite{Klebanov:1999tb}.  For odd $d$, our Neumann theory is
conformally- (in fact, Weyl-) invariant and is much like the Neumann
theory for a scalar with $m_{BF}^2 +1 > m^2 > m_{BF}^2$, while for
even $d$ it is much like the logarithmic Neumann theory associated
with scalars that saturate the Breitenlohner-Freedman bound ($m^2 =
m_{BF}^2$).  We will show in particular that for odd $d$ our Neumann
boundary condition $T_{ij} =0$ leads to a conformal gravity theory,
in which both diffeomorphisms and Weyl rescalings of $g_{(0)ij}$ are
gauge transformations.  For such cases, it is reasonable to
expect the theory to be UV complete.  Moreover, for such odd $d$ we show that the spectrum of the boundary graviton is ghost-free around flat space; at least at this level, the theory is unitary.  We do find ghosts for even $d$, though for  $d = 2$ the boundary theory may be recognized as (timelike) Liouville theory coupled to the original CFT and to a fluctuating metric; i.e., as a supercritical string theory on the worldsheet.

Similar results also hold for much more general boundary conditions.
Let us return to the path-integral argument above, this time
coupling the CFT to a $d$-dimensional gravity theory defined by some
action $S_{Bndy \ grav}[g_{(0)}]$ (which may include cosmological
terms, Einstein-Hilbert terms, and higher derivative terms).  One
may regard this as merely a deformation of the (Neumann) induced
gravity theory, and the deformation is relevant so long as $S_{Bndy
\ grav}[g_{(0)}]$ contains only operators of dimension $\le d$.  In
this case one expects the asymptotic behavior of the bulk metric to
approach that of the Neumann theory, and in particular that the
Fefferman-Graham expansion (\ref{eq:FG1}) remains valid. Thus, the
deformation merely inserts a finite factor of $\exp(-S_{Bndy \
grav}[g_{(0)}])$ into the bulk path integral, adding $S_{Bndy \
grav}[g_{(0)}]$ to $S_{Neu}$. The result is an AdS bulk theory
associated with the `mixed' (Dirichlet-Neumann) boundary condition
\begin{equation}
\label{mixed} \frac{2}{\sqrt{-g_{(0)}}} \frac{\delta S_{Bndy \
grav}}{\delta g_{(0)ij}} +  T^{ij} = 0.
\end{equation}
We will show below that fluctuations satisfying (\ref{mixed})
conserve the (finite) symplectic structure defined by $S_{total} =
S_{Bndy \ grav} + S_{Neu}$.

The plan of our paper is as follows.  Section 2 describes the
contributions of the counter-terms to the symplectic structure,
shows that the total symplectic structure is conserved under the
boundary conditions (\ref{mixed}), and verifies that fluctuations
 have finite norm for $d=2,3,4$.  Section 3
verifies that the bulk AdS theory has the properties one would
expect of a gravitational theory of $g_{(0)}$, namely that
$d$-dimensional diffeomorphisms and (for odd $d$ and appropriate
$S_{Bndy \ grav}$) that Weyl rescalings are gauge symmetries.  In
particular, taking the AdS boundary to be $\tilde K \times {\mathbb
R}$ for any compact manifold $\tilde K$ causes the conserved charge
associated with any such transformation to vanish.  We then briefly
examine the dynamics of some of these theories by studying
linearized fluctuations about AdS${}_{d+1}$ and computing the
boundary graviton propagator.  The $d=2$ case is known to give
Liouville gravity, and we study the Neumann theories for $d \ge 3$
in section \ref{dynamics}.  In the perturbative expansion around
flat space, these theories are ghost- and tachyon- free for odd $d$,
while they contain both ghosts and tachyons for even $d$.  For
$d=3$, we also consider deformations by terms associated with
topologically massive gravity (TMG).  We find that coupling TMG to a
large $N$ CFT preserves the perturbative stability of the theory
with negative (3-dimensional) Newton's constant.

\section{Counterterms and the symplectic structure}
\label{sec:CTS}

We wish to show that contributions from boundary counter-terms in
the bulk gravitational action render normalizable all fluctuations
of the bulk metric $g_{ij}$.  We begin with a brief review of
counter-term technology and the construction of the desired
symplectic structure in section \ref{sec:extendedsymplectic}.  We
then verify that this symplectic structure is conserved (section
\ref{conserved}) under the appropriate boundary conditions and
(section \ref{finite}) that all fluctuations are normalizable.

\subsection{Counterterms and the symplectic structure}
\label{sec:extendedsymplectic}

We wish to investigate the normalizability of certain modes in a
given theory of AdS gravity.  To do so, it is critical to begin with
the correct inner product.  For familiar scalar fields, this is
simply the Klein-Gordon inner product.  More generally, however,
this is given by the symplectic structure of the theory, which is
closely associated with the commutation relations (see e.g.
\cite{WaldThermo}).

Let us now consider the action $S_{Neu}$ in a context where
$g_{(0)ij}$ is free to vary.   Since the counter-terms generally
contain derivatives of $g_{(0)ij}$, it is clear that the
counter-terms contribute to the definition of the canonical momenta
and thus to the commutation relations and symplectic structure.
They should therefore affect the condition for normalizeability as
well.

Instead of proceeding via canonical methods, we find it more
convenient to take a covariant phase space approach. The results
should be equivalent by the usual arguments.  Recall that the usual
Einstein-Hilbert symplectic structure is defined via the observation
that the variation of the Einstein-Hilbert action reduces on-shell
to a boundary term $\delta L_{EH} \approx \dH\Theta_{EH}[\delta g]$,
where $\Theta_{EH}[\delta g]$ defines the pre-symplectic structure.
The symplectic structure is then\footnote{Here we have defined
$\delta \equiv \delta g_{\alpha\beta}\frac{\d}{\d
g_{\alpha\beta}}+\d_{\mu}\delta g_{\alpha\beta}\frac{\d}{\d
\d_{\mu}g_{\alpha\beta}}+\cdots$ as an operator with $\delta g$
Glassmann even and $\omega_{EH}[\delta g,\delta g]$ is a two-form in
variations $\delta g$, see \cite{Compere:2007az,Barnich:2007bf} for
details. Smearing this two-form with particular variations $\delta_1
g$ and $\delta_2 g$, we get $\omega[\delta_1 g,\delta_2 g]= \delta_1
\Theta[\delta_2 g] - (1 \leftrightarrow 2)$.} $\omega_{EH}[\delta
g,\delta g] \equiv \delta \Theta_{EH}[\delta g]$.  This prescription
admits the well-known ambiguity $\Theta \rightarrow \Theta + \dH B$
\cite{Iyer:1994ys,Wald:1999wa}. For definiteness we take
$\Theta_{EH}[\delta g]$ to be the standard pre-symplectic structure
given in \cite{Wald:1999wa}, about which we will say more in section
\ref{finite}.  Below, we will argue that considering the full action
including all boundary terms leads to a preferred choice for $B$.

To do so, we first recall (see e.g. \cite{Papadimitriou:2005ii})
that the counter-terms are defined by considering similar variations
of the action. Let us write the full action $S_{Neu}$ as
\begin{equation}
S_{Neu} = \int_\cM  L_{EH} + \int_{\partial \cM} L_{GH} +
\int_{\partial \cM} L_{ct}\label{actionS},
\end{equation}
with $L_{EH} = \frac{1}{16\pi G} \sqrt{-g}(R-2\Lambda) d^{d+1}x$,
$L_{GH} = \frac{1}{8\pi G}\sqrt{-\gamma}K d^d x$, where
$\gamma_{ij}$ was introduced in (\ref{eq:FG1}) and where we suppose
for simplicity that the only boundary $\partial \cM$ is spatial
infinity, ignoring any boundary terms that may arise at past or
future boundaries. We use here the Wheeler conventions for the
Riemann tensor so that AdS has negative curvature $R =
+\frac{2(d+1)}{d-1}\Lambda = -d(d+1)/\ell^2$. The extrinsic
curvature is defined as $K = \gamma^{ij}\nabla_i n_j$ where $n^i$ is
the outward pointing unit normal and $\int_\cM d^{d+1}x = \int_\cM
d^{d}x \int^\infty_0 dr$.

The action \eqref{actionS} is defined by first introducing a cut-off
at $r^2 = \eps$ and then taking the limit $\eps \rightarrow 0$. This
of course requires choosing a specific radial foliation close to the
boundary $\d \cM$, which specifies a Fefferman-Graham coordinate
system. For odd dimensions $d$, the total action $S_{Neu}$ turns out
not to depend on the choice of radial foliation.  This is precisely
the statement that the holographic conformal anomaly vanishes for
odd $d$. However, for even $d$ the choice of radial foliation does
affect the definition of the counter-terms.

The counter-terms are defined by the property that all divergent
boundary terms cancel in the variation $\delta S_{Neu}$; i.e., by
the condition
 \begin{equation}
 \label{ctDef} \Theta_{EH}[\delta g]|_{\partial \cM} + \delta L_{GH}
 \approx - \delta \gamma_{ij} \varQ{L_{ct}}{\gamma_{ij}} +
 \frac{1}{2}\sqrt{-g_{(0)}} T^{ij} \delta g_{(0)ij}d^d x,
 \end{equation}
where $T^{ij}$ is finite ($T^{ij} = O(\eps^0)$), $\delta \gamma_{ij}
\varQ{L_{ct}}{\gamma_{ij}} = \delta L_{ct}[\gamma] - \dH
\Theta_{ct}[\delta \gamma]$ and $\dH$ is the induced $d$-dimensional
total derivative on hypersurfaces $r=$ constant. Here we understand
$g_{(0)ij}$ as a function of $\gamma_{ij}$, so that any dependence
on $g_{(0)ij}$ contributes to the Euler-Lagrange derivative on the
right-hand side.  The boundary stress tensor is covariantly
conserved on-shell ($D_j T^{ij} \approx 0$) with respect to the
covariant derivative $D$ compatible with $g_{(0)ij}$: It has a trace
which vanishes for $d$ odd and which is related to the conformal
anomaly in even $d$. When Einstein's equations hold, the entire F-G
expansion is determined in terms of $g_{(0)ij}$ and $T^{ij}$.
Throughout this work the symbol $\approx$ expresses equality up to
terms vanishing when the equations of motion hold and, when
linearized equations are involved, up to terms vanishing when the
linearized equations of motion hold as well.

The boundary terms $\Theta_{ct}$ are usually discarded as they
contribute only at the boundary of the surface $r^2 =\eps$; i.e., at
boundaries of the spacetime describing the dual field theory. We
observe, however, that since (\ref{ctDef}) requires $\Theta_{ct}
\rightarrow \Theta_{ct} + B$, the combination $\Theta_{EH}[\delta
g]|_{\partial \cM} - \dH \Theta_{ct}[\delta \gamma]$ is invariant
under any shift $\Theta_{EH} \rightarrow \Theta_{EH} + \dH B$
associated with the ambiguity in the pre-symplectic structure. This
motivates us to introduce the symplectic structure
 \begin{eqnarray} \omega_{Neu}[\delta g,\delta g] =
 \omega_{EH}[\delta g,\delta g] - \dH \omega_{ct},
 \label{eq:defomegaform}\label{def:Omega}
 \end{eqnarray}
which is similarly invariant under $\omega_{EH} \rightarrow
\omega_{EH} + \dH\delta B$.  In effect, we have fixed the ambiguity
by appealing to the boundary terms in the action.  In
\eqref{def:Omega} we have chosen an arbitrary smooth extension of
$\Theta_{ct}$ away from the boundary and defined $\omega_{ct} \equiv
\delta \Theta_{ct}$ by varying both $\gamma$ and $g_{(0)}$.

Our main claim, to be detailed in the following sections, is that
the symplectic structure $\Omega_{Neu,\Sigma} = \int_\Sigma
\omega_{Neu}$ is finite for arbitrary variations respecting the
Fefferman-Graham form, and that it is conserved when one imposes
Neumann boundary conditions.  It is clear that the addition of
further boundary terms $S_{Bndy \ grav}$ to $S_{Neu}$ leads by an
analogous construction to additional finite boundary terms in the
symplectic structure, and we show that the result is conserved under
the boundary conditions \eqref{mixed}.

\subsection{Conservation of the symplectic structure}
\label{conserved}

Consider a spacetime volume $\cV$ whose boundary consists of two
spacelike surfaces $\Sigma_1$, $\Sigma_2$ and the timelike surface
$\cT \subset \partial M$ at spatial infinity, $\partial \cV =
\Sigma_2 - \Sigma_1
 -\cT$. Since the usual symplectic structure is conserved on-shell, $\dH \omega_{EH} \approx 0$, We have
\begin{equation}
\Omega_{Neu,\Sigma_2} - \Omega_{Neu,\Sigma_1} \approx \int_{\cT}
\omega_{EH}[\delta g,\delta g] - \int_{\partial \cT}
\omega_{ct}[\delta\gamma,\delta \gamma].\label{eq:10}
\end{equation}
Now recall that the counterterms were defined (\ref{ctDef}) so that
 $\Theta_{EH}|_{\cT}$ reduces on-shell
to exact variations $\delta L_{GH} + \delta L_{ct}$ (which do not
contribute to the symplectic structure $\omega_{EH}|_{\cT}$), to the
boundary term $\dH \Theta_{ct}$ (which cancels exactly the boundary
contribution in \eqref{eq:10}) and to a finite contribution which
defines the boundary stress tensor $T^{ij}$. In summary,
\begin{eqnarray}
\Omega_{Neu,\Sigma_2} - \Omega_{Neu,\Sigma_1} &\approx & \int_{\cT}
d^dx \frac{1}{2}\delta (\sqrt{-g_{(0)}} T^{ij}) \delta g_{(0)ij}
.\label{eq:conservation}
\end{eqnarray}
It follows that the symplectic structure is independent of $\Sigma$
if one of the following conditions holds:
\begin{enumerate}
\item Dirichlet boundary conditions:
 \begin{equation}
 g_{(0)ij} = \bar g_{(0)ij}.
 \end{equation}
Fixing the conformal representative $g_{(0)}$ to a prescribed value
breaks conformal invariance and introduces a background structure at
the boundary. For any $d$, under such boundary conditions we have
$\omega_{Neu} = \omega_{EH}$.

\item Neumann boundary conditions:
\begin{equation}
T^{ij} = 0.
\end{equation}
In particular, for $d$ even the equations of motion will impose that
$g_{(0)}$ has no trace anomaly. It is crucial for this boundary
condition that the symplectic structure be defined by
\eqref{def:Omega}.  In particular,  $\Omega_{EH,\Sigma} =
\int_\Sigma \omega_{EH}$ is not conserved\footnote{As remarked in
\cite{Papadimitriou:2005ii}, one also obtains a valid variational
principle by imposing the boundary conditions
 $g_{(0)ij} = e^{2 \sigma} \bar g_{(0)ij}$, $T^i_{\;i}[g_{(0)}] =
 0$,
with $\sigma$ free to vary but $\bar g_{(0)ij}$ held fixed. For this
case (\ref{def:Omega}) again yields a conserved  symplectic
structure.}.

\item Mixed boundary conditions:

If $S_{Bndy \ grav}$ is a finite action built only from $g_{(0)}$
(and perhaps certain non-dynamical background structures), then the
action $S_{total} = S_{Neu} + S_{Bndy \ grav}$ satisfies the
analogue of (\ref{ctDef}) but with $T^{ij}$ replaced by
 \begin{equation}
 T_{total}^{ij} \equiv  T^{ij}  + \frac{2}{\sqrt{-g_{(0)}}}\varQ{L_{Bndy \ grav}
 [g_{(0)}]}{g_{(0)ij}},\label{totalstress1}
 \end{equation}
 and with $L_{ct}$ replaced by $L_{ct} + L_{Bndy\ grav}.$
As a result, the argument above shows that the total symplectic
structure
\begin{eqnarray}
\omega_{total}[\delta g,\delta g] &\equiv & \omega_{EH}[\delta
g,\delta g] - \dH \omega_{ct} - \dH \omega_{Bndy \
grav}.\label{eq:OmegaB}
\end{eqnarray}
is conserved when $T_{total}^{ij}=0$ (i.e., when (\ref{mixed})
holds). Here
 $\omega_{Bndy \
grav}[\delta g_{(0)},\delta g_{(0)}] \equiv \delta \Theta_{Bndy \
grav}[\delta g_{(0)}]$ and $\Theta_{Bndy \ grav}[\delta
g_{(0)},\delta g_{(0)}]$ is the boundary term relating the
Euler-Lagrange derivatives $\varQ{L_{Bndy \ grav}}{g_{(0)ij}}$ to
the variations $\delta L_{Bndy \ grav}$.

\end{enumerate}

Since for each boundary condition the associated symplectic
structure is independent on $\Sigma$, we will denote it by simply
$\Omega_{total}$. In the notation of \eqref{eq:defomegaform}, the
boundary conditions impose in each case that
$\omega_{total}|_{\partial M} = 0$.

\subsection{Finiteness}
\label{finite}

We are interested in the conserved symplectic structure
$\Omega_{total}$.  However, for any boundary condition,
$\Omega_{total}$ differs from $\Omega_{Neu}$ only by finite terms.
Show that $\Omega_{Neu}$ is finite for arbitrary variations of the
metric thus implies the same result for $\Omega_{total}$. We verify
below that this is the case for $d \leq 4$ by using the explicit
form of the counterterms, but we expect the same result to hold for
higher $d$ as well.

The counterterm action is given by \cite{deHaro:2000xn}
 \begin{eqnarray}
 S_{ct} = -\frac{1}{16\pi G} \int d^d x \left[ 2(d-1)\sqrt{-\gamma}  + \frac{1}{d-2}\sqrt{-\gamma} R[\gamma] + \cdots  +  \log{\eps} A_{(d)}\right]_{r^2 =
 \eps}\label{ctaction},
 \end{eqnarray}
where the second term arises only for $d\geq 3$ and the dots
indicate higher curvature terms for $d\geq 5$. The logarithmic
divergence appears only for $d$ even,
\begin{eqnarray}
A_{(2)} = \sqrt{-g_{(0)}}g_{(2)i}^{\;\;\, i},\qquad A_{(4)} = \half \sqrt{-g_{(0)}} ( g_{(2)i}^{\;\;\, i} g_{(2)j}^{\;\;\, j} - g_{(2)i}^{\;\;\, j} g_{(2)j}^{\;\;\, i}), \;\dots
\end{eqnarray}
or, when Einstein's equations hold,
\begin{eqnarray}
A_{(2)} = -\frac{1}{2} \sqrt{-g_{(0)}}R_{(0)},\qquad A_{(4)} = -\frac{1}{8} \sqrt{-g_{(0)}}( R^{ij}_{(0)}R_{(0)ij} - \frac{1}{3}R^2_{(0)}), \;\dots \label{eq:anomal}
\end{eqnarray}
Viewing the counterterm Lagrangian as an action for $\gamma$ and
$g_{(0)}$, one may calculate its symplectic structure $\omega_{ct}$,
\begin{eqnarray}
\omega_{ct} = -\frac{1}{d-2}\omega^{(d)}_{EH}[\delta \gamma,\delta
\gamma] +\cdots -\frac 1 {16\pi G} \log{\eps}
\;\omega_{A_{(d)}}[\delta g_{(0)},\delta g_{(0)}],\label{eq:omegact}
\end{eqnarray}
where $\omega^{(d)}_{EH}[\delta \gamma, \delta \gamma]$ denotes the
usual symplectic structure of the $d$-dimensional Einstein-Hilbert
action $S^{(d)}_{EH} = 1/(16\pi G)\int d^d x \sqrt{-\gamma}
R[\gamma]$ and $\omega_{\frac{1}{16\pi G} A_{(2)}} = -
\frac{1}{2}\omega_{EH}[\delta g_{(0)},\delta g_{(0)}]$. The
symplectic structure $\omega_{\frac{1}{16\pi G} A_{(4)}}$ is
computed in Appendix~\ref{app:Weylomega}.

Let us now consider the bulk symplectic form,
\begin{equation}
\label{OmegaEps} \omega_{EH}[\delta g,\delta g] = \frac{1}{d!}\omega^\mu_{EH}
\epsilon_{\mu \alpha^1 \cdots \alpha^d}dx^{\alpha^1}\cdots
dx^{\alpha^d}.
\end{equation}
An explicit expression can be found e.g. in \cite{Wald:1999wa}. It
was shown in \cite{Barnich:2007bf} that $\omega_{EH}$ can be written
as minus the so-called invariant symplectic form, defined
algebraically from the equations of motion, and a certain boundary
term $E_\cL$. When the metric and its perturbation preserve the
Fefferman-Graham form to the order in $r$ stated in (\ref{eq:FG2}),
$E_\cL$ vanishes identically when integrated over a closed surface
at fixed time and radius. We will thus ignore the term $E_\cL$ and
write the symplectic form as
\begin{eqnarray}
\omega^\mu_{EH}[\delta_1 g,\delta_2 g] =
- P^{\mu\nu\beta\gamma\varepsilon\zeta}\Big( \delta_2 g_{\beta\gamma}\nabla_\nu \delta_1 g_{\varepsilon\zeta}- (1 \leftrightarrow 2) \Big),
\label{omegaEHgamma}
\end{eqnarray}
where
\begin{eqnarray}
P^{\mu\nu\alpha\beta\gamma \delta}  = \frac{\d}{\d g_{\gamma
\delta,\alpha\beta}} \left( \varQ{L_{EH}}{g_{\mu\nu}} \right)
\end{eqnarray}
and $g_{\gamma \delta,\alpha\beta}$ are second derivatives of
$g_{\gamma \delta}$. Explicitly, we have
\begin{eqnarray}
P^{\mu \nu \alpha\beta \gamma\delta} &= &\frac{\sqrt{-g}}{32\pi G}
\big( g^{\mu\nu} g^{\gamma(\alpha} g^{\beta)\delta} +
g^{\mu(\gamma}g^{\delta)\nu} g^{\alpha\beta} + g^{\mu(\alpha}
g^{\beta)\nu}g^{\gamma\delta}
\nonumber \\
&& - g^{\mu\nu} g^{\gamma\delta} g^{\alpha\beta}  -
g^{\mu(\gamma}g^{\delta)(\alpha}g^{\beta)\nu}  -g^{\mu(\alpha}
g^{\beta)(\gamma}g^{\delta)\nu} \big).\label{tensorP}
\end{eqnarray}
The tensor density $P^{\mu\nu\alpha\beta\gamma \delta}$ is symmetric
in the pairs of indices $\mu\nu$, $\alpha\beta$, and $\gamma \delta$
and the total symmetrization of any three indices is zero.
See~\cite{Barnich:2007bf} for additional comments.

Having imposed the Fefferman-Graham gauge choice (\ref{eq:FG1}), it
turns out that the pull-back of the symplectic form
$\omega_{EH}[\delta g,\delta g]$ to $\Sigma$ can be expressed in
terms of the $d$-dimensional Einstein-Hilbert symplectic structure
$\omega^{(d)}_{EH}[\delta \gamma,\delta \gamma]$.  The point is that
\eqref{eq:FG1} fixes the radial components of $g_{\mu \nu}$ so that
variations $\delta g_{\mu \nu}$ reduce to variations $\delta
\gamma_{ij}$.  Choosing coordinates such that $t=$ constant on
$\Sigma$, we see from (\ref{OmegaEps}), (\ref{tensorP}) that the
pull-back of $\omega^t_{EH}[\delta g,\delta g]$ contains no radial
derivatives.  As a result, we have
\begin{eqnarray}
\omega^{t}_{EH}[\delta g,\delta g] = \frac{1}{r}
\omega^{(d)t}_{EH}[\delta \gamma,\delta \gamma],
\end{eqnarray}
so that
\begin{eqnarray}
\Omega_{Neu} = \int d\Omega_{d-1} \left( \int_{\sqrt\epsilon}
\frac{dr}{r} \omega^{(d)t}_{EH}[\delta \gamma,\delta \gamma] +
\omega^t_{ct}|_{r=\sqrt{\eps}}\right),
\end{eqnarray}
where $d\Omega_{d-1}$ denotes the volume form on a sphere of
constant $r$ and $t$. We have suppressed the upper limit of
integration for $r$ as it is not relevant for studying
normalizeability of fluctuations near $r=0$. Using the
Fefferman-Graham expansion on-shell $\gamma \approx r^{-2}
g_{(0)}+g_{(2)}[g_{(0)}]+O(r)$, one finds
\begin{eqnarray}
\omega^{(d)}_{EH}[\delta \gamma,\delta \gamma] \approx r^{2-d}\omega_{(0)EH}[\delta g_{(0)},\delta g_{(0)}]+r^{4-d}\omega_{(2)EH}[\delta g_{(0)},\delta g_{(0),(2)}]+O(r^{5-d}),\label{eq:omegaEHgamma}
\end{eqnarray}
where $\omega_{(2)EH}[\delta g_{(0)},\delta g_{(0),(2)}]$ is a $d$-form which depends on $\delta g_{(0)}$ and either $\delta g_{(0)}$ or $\delta g_{(2)}$. In four dimensions, it is shown in Appendix~\ref{app:Weylomega} that $\omega_{(2)EH}[\delta g_{(0)},\delta g_{(0),(2)}] =
-\frac{1}{8\pi G} \omega_{A_{(4)}}[\delta g_{(0)},\delta g_{(0)}]$ when $g_{(2)}$ and its variation are expressed in terms of
$g_{(0)}$.

Inserting expression ~\eqref{eq:omegact} for $\omega_{ct}$  and
expanding as in \eqref{eq:omegaEHgamma}, one finds explicit
cancelation of the divergent terms for $2\leq d\leq 4$.     In the
notation of \eqref{eq:defomegaform}, we have shown $\omega^t_{Neu} =
O(r^0)$ for arbitrary variations $\delta g$. Thus the symplectic
structure is finite as claimed.  One expects this result to hold for
$d \ge 5$ as well, though we have not analyzed such cases in detail.

\section{Gauge Transformations and Conserved charges}

Boundary conditions play an important role in gravitational
theories.  They determine precisely which diffeomorphisms are gauge
transformations, which diffeomorphisms act as global symmetries
generated by non-trivial charges, and which diffeomorphisms  do not
preserve the phase space at all.  We investigate this issue below by
considering the conserved charges associated with all
diffeomorphisms that preserve a given boundary condition. Our goal
is to show that the bulk system has the properties one would expect
from a gravitating theory on the boundary.  Namely, for Neumann or
mixed boundary conditions where $S_{Bndy \ grav}$ is built only from
$g_{(0)ij}$, we show that all diffeomorphisms of the boundary are
gauge transformations.  We also show that boundary Weyl
transformations are pure gauge for odd $d$ and Weyl-invariant
$S_{Bndy \ grav}$.

We begin by reviewing the construction of conserved charges for
Dirichlet boundary conditions and then proceed to the cases of
Neumann and mixed boundary conditions. In the rest of this section
we take the boundary manifold $\partial M = \partial \Sigma \times
{\mathbb R}$ to have compact spatial sections $\partial \Sigma$.  We
also largely follow the notation and definitions of
\cite{Barnich:2001jy,Compere:2007az,Barnich:2007bf}; see in particular appendix A
of \cite{Compere:2007az}.

\subsection{Dirichlet problem}

An infinitesimal diffeomorphism can only be a symmetry or gauge
transformation of a given system if the diffeomorphism maps the
phase space to itself. We refer to such diffeomorphisms as `allowed'
for that system.   Since we have required the leading terms in the
metric to take the Fefferman-Graham form (\ref{eq:FG1}), such
transformations must leave this form invariant. Recall that this
occurs precisely~\cite{Papadimitriou:2005ii}
 for diffeomorphisms deep in the bulk, i.e. such
that $\cL_\xi g_{ij} = O(r^{d-1})$, $\cL_\xi g_{ri} = O(r^{d})$,
$\cL_\xi g_{rr} = O(r^{d+1})$,  and for the additional
diffeomorphisms
\begin{eqnarray}
\label{BdiffW}
\xi^r &=& r \delta \sigma(x^i),\label{eq:conf}\\
\xi^i &=& \phi^i(x^j) - \d_j\delta \sigma(x^k) \int^r
\frac{dr^\prime}{r^\prime} l^2 \gamma^{ij}(r^\prime)\nonumber.
\end{eqnarray}
The latter induce a Weyl transformation with conformal factor
$\delta \sigma(x^i)$ at the boundary and a boundary diffeomorphism
generated by $\phi^i(x^j)$. In particular, the induced
transformation of $g_{(0)}$ (and, for later use, of the
stress-tensor) read
\begin{eqnarray}
\delta g_{(0)ij} &= & 2 \delta \sigma g_{(0)ij}+\cL_{\phi^{(i)}} g_{(0)ij},\label{eq:34}\\
\delta T_{ij} &=& -(d-2) \delta \sigma T_{ij}+(\text{anomalous
term})+\cL_{\phi^{(i)}} T_{ij}.\label{eq:35}
\end{eqnarray}

For Dirichlet boundary conditions we are thus lead to demand that
$\delta g_{(0)ij} =  2 \delta \sigma g_{(0)ij}+\cL_{\phi^{(i)}}
g_{(0)ij} = 0$, so that the allowed set of diffeomorphisms is
spanned by the conformal Killing vectors of $g_{(0)}$ together with
diffeomorphisms deep in the bulk.  The infinitesimal charge
difference between two solutions $g$ and $g+\delta g$ associated to
a (conformal) Killing vector of the boundary metric may be
constructed by integrating the $(d-1)$-form\footnote{Our definition
agrees with \cite{Iyer:1994ys,Wald:1999wa} but has an extra minus
(-) sign relative to \cite{Compere:2007az,Barnich:2007bf}.}
\begin{equation}
k^{EH}_\xi[\delta g] \equiv -I_\xi \omega_{EH}[\cL_\xi g,\delta g],\label{eq:linchargeD}
\end{equation}
over the boundary, where $I_\xi$ is a particular homotopy operator
(see \cite{Compere:2007az,Barnich:2007bf} and the original
references
\cite{Andersonbook,Olver:1993,Saunders:1989,Dickey:1991xa}). The
main property we will use here is that, when acting on a $p$-form
linear in $\xi$ and its derivatives, $I_\xi$ satisfies $dI_\xi +
I_\xi d = 1$. Evaluating the charge on the $(d-1)$-surface $\partial
\Sigma$, one can write
\begin{equation}
\oint_{\partial \Sigma} k^{EH}_\xi[\delta g] = \delta
\oint_{\partial \Sigma} k^K_\xi - \oint_{\partial \Sigma} i_\xi
\Theta_{EH}[\delta g],\label{eq:originalEH}
\end{equation}
where $\oint_{\partial \Sigma} k^K_\xi$ is the Komar charge defined
in \cite{Iyer:1994ys,Wald:1999wa}, $\Theta_{EH}[\delta g]$ is the
presymplectic form, and $i_\xi$ is the interior product\footnote{As
noted previously, the additional term involving $E_\cL[\delta
g,\delta g]$ in \cite{Compere:2007az,Barnich:2007bf} vanishes here
due to the Fefferman-Graham form of the metric.}. These
infinitesimal charges are conserved, finite and integrable for
variations preserving (\ref{eq:FG1}) \cite{Papadimitriou:2005ii}.
Thus they may be expressed as exact variations, $k_\xi[\delta g] =
\delta K^D_\xi$. The integrated charge $\int_{\partial \Sigma}
K^D_\xi$ is equal on-shell, up to an undetermined constant offset,
to the charge defined by the counterterm procedure
\cite{Hollands:2005wt,Hollands:2005ya}
\begin{equation}
\cQ^D_\xi[T_{ij}] \approx \oint_{\partial \Sigma}\left( \sqrt{-g_{(0)}}  T_{ij} n^i[ g_{(0)}] \xi^j \right).\label{eq:chargeD}
\end{equation}

\subsection{Neumann problem}

In the Neumann case, we will assume that $\partial \Sigma := \Sigma
\cap \partial M$ is a compact manifold without boundary.  As a
result, for odd $d$ all bulk diffeomorphisms of the
form~\eqref{eq:conf} preserve the phase space: arbitrary Weyl
transformations and diffeomorphisms of the boundary are allowed, as
well as diffeomorphisms deep in the bulk.  However, due to the
anomalous term in \eqref{eq:35}, Weyl transformations do not
preserve $T_{ij}=0$ when $d$ is even.  In this case the allowed
diffeomorphisms are only those with $\delta \sigma =0$.

For the associated vector fields $\xi$, the Neumann symplectic
structure defines the infinitesimal charges
\begin{equation}
k_\xi[\delta g]\equiv -I_\xi \omega_{Neu}[\cL_\xi g,\delta g] =
-I_\xi \omega_{EH}[\cL_\xi g,\delta g] +\omega_{ct}[\cL_\xi g,\delta
g].
\end{equation}
It is easy to prove that these charges are conserved when both the
equations of motion and the linearized equations of motion hold.
Indeed, we have
\begin{eqnarray}
\dH k_\xi &=& -\omega_{Neu}[\cL_\xi g,\delta g]+I_\xi \dH
\omega_{Neu}[\cL_\xi g,\delta g]\\
 &=& -\omega_{Neu}[\cL_\xi g,\delta g]+I_\xi (\cL_\xi g_{\mu\nu} \delta
\varQ{L_{EH}}{g_{\mu\nu}}-\delta g_{\mu\nu} \delta_\xi \varQ{L_{EH}}{g_{\mu\nu}})\\
 &\approx &-\omega_{Neu}[\cL_\xi g,\delta g].
\end{eqnarray}
Conservation follows from the boundary condition $\omega_{Neu}|_{\d
\cM} = 0$, and finiteness of the charges follows from the finiteness
of the symplectic structure.

The charges are also integrable. The proof is instructive.
\begin{eqnarray}
\delta k_\xi &=& -i_\xi \omega_{EH} + \delta i_{\cL_\xi g} \omega_{ct}\\
&=& -i_\xi \omega_{Neu} + (\{\delta ,i_{\cL_\xi g}\} - \{ i_\xi,\dH
\}
)\omega_{ct}+\dH(\cdot)\\
&=& -i_\xi \omega_{Neu} + (\delta_{\cL_\xi g} - \{ i_\xi,\dH \}
)\omega_{ct}+\dH(\cdot),
\end{eqnarray}
where $\dH(\cdot)$ is an exact form that we need not write
explicitly. Now, because $\omega_{ct}$ does not depend on background
structures, diffeomorphism invariance requires the term proportional
to $\omega_{ct}$ to vanish. Integrating on the sphere at infinity
one finds $\delta \int_{S^\infty} k_\xi  = 0$ from the conservation
condition $\omega_{Neu}|_{\d \cM} = 0$, the finiteness result
$\omega^t_{Neu} = O(r^0)$, and from the fact that diffeomorphisms
\eqref{eq:conf} satisfy $\xi^t = O(r^0)$, $\xi^r = O(r)$ so that
both $\xi^r \omega^t_{Neu}$ and $\xi^t \omega^r_{Neu}$ vanish on
$\partial M$. Due to the definition of $\delta$
\cite{Compere:2007az,Barnich:2007bf}, this implies that
$\int_{S^\infty} k_\xi$ is an exact variation.

It follows that each boundary diffeomorphism (and, for $d$ odd, each
boundary Weyl transformation) leads to a finite conserved charge.
However, because one is free to vary $\xi$ locally on the boundary,
all of these charges must vanish. Indeed, one may choose a vector
field $\xi$ that is non-zero only in a compact region of the
boundary. In the region where $\xi=0$, the charge is manifestly
zero.  Thus by conservation it must vanish everywhere. Since such
vector fields span the space of all vector fields, all charges
vanish on-shell.

Furthermore, since $\Omega_{Neu}[\cL_\xi g, \delta g]$ is the
variation of such a (vanishing) charge, each allowed diffeomorphism
$\xi$ gives a degenerate direction of the symplectic structure. We
see that every allowed bulk diffeomorphism is a gauge
transformation. From the boundary perspective, one finds from
(\ref{BdiffW}) that boundary diffeomorphisms and (for odd $d$)
boundary Weyl rescalings are gauge transformations. Boundary
conditions with similar properties for finite boundaries were
explored in e.g. \cite{SmolinAction,Smolin:2003qu}.

We can use these results to write an explicit formula for the
charges corresponding to diffeomorphisms at the boundary ($\xi^r=0$,
$\xi^i = \phi^i(x^j)$) which will be useful for general boundary
conditions below. For such $\xi$, $\oint_{S^\infty} i_\xi
\Theta[\delta g]$ depends only on the pull-back of $\Theta[\delta
g]$ to $\partial M$ and one can use the counterterm recipe
(\ref{ctDef}) to show that for any boundary condition we have
\begin{equation}
\oint_{S^\infty} k_\xi[\delta g] \approx  \delta \oint_{S^\infty}
(k^K_\xi +i_\xi L_{GH}+i_\xi L_{ct}-\Theta_{ct}[\cL_{\xi}\gamma]) -
\oint_{S^\infty} i_\xi \Theta_{(0)}[\delta g_{(0)}], \label{prech}
\end{equation}
where $\Theta_{(0)}[\delta g_{(0)}]= \frac{1}{2}
\sqrt{-g_{(0)}}T^{ij}\delta g_{(0)ij} d^d x$. This formula is very
similar to the expression for the original Einstein-Hilbert charge
\eqref{eq:originalEH}, but with the Komar term supplemented by
additional terms implied by the regularization procedure and with
the presymplectic contribution simplified to $\Theta_{(0)}[\delta
g_{(0)}]$.  This $\Theta_{(0)}[\delta g_{(0)}]$ can be called the
finite presymplectic form for $g_{(0)}$.

Now, for both Dirichlet and Neumann boundary conditions the term
$\Theta_{(0)}[\delta g_{(0)}]= \frac{1}{2}
\sqrt{-g_{(0)}}T^{ij}\delta g_{(0)ij} d^d x$ vanishes. In the
Dirichlet case, it is known that any difference between charges
obtained by covariant space phase methods and those obtained by
counterterm methods must be a function only of the background metric
$g_{(0)}$ \cite{Hollands:2005wt,Hollands:2005ya}. There is a charge
for every Killing field, and for any boundary vector field $\xi^j$
there is some metric for which $\xi^j$ is a Killing field. Thus the
exact term in \eqref{prech} must in general be
 the variation of \eqref{eq:chargeD} up to a term $\delta
\oint_{S^\infty} F_{ij}[g_{(0)}]n^i[ g_{(0)}] \xi^j $ depending
only\footnote{The above logic also allows a general term which
depends on $T_{ij}$ and contains a factor of ${\cal L}_\xi
g_{(0)ij}$. However, such terms have the wrong structure to arise
from the Komar or $L_{GH}$ terms, and counterterm contributions do
not involve $T_{ij}$.} on $g_{(0)}$. On the other hand, with Neumann
boundary conditions we proved that all charges vanish. Since
\eqref{eq:chargeD} and $\Theta_{(0)}[\delta g_{(0)}]$ both zero, it
follows that $\delta \oint_{S^\infty} F_{ij}[g_{(0)}]n^i[ g_{(0)}]
\xi^j$ also vanishes.  We conclude that for any boundary conditions
and for any allowed $\xi$ the variation of the
 associated charge is given by
\begin{equation}
\oint_{S^\infty} k_\xi[\delta g] \approx \delta \oint_{S^\infty}
\left( \sqrt{-g_{(0)}}  T_{ij} n^i[ g_{(0)}] \xi^j \right)
-\oint_{S^\infty} i_\xi \Theta_{(0)}[\delta g_{(0)}].
\label{chargeLB}
\end{equation}
We have also checked this formula explicitly for $d=3$.

\subsection{Mixed problem}

Finally, let us suppose that the action is supplemented by an extra
boundary term $S_{Bndy \  grav}$ $= \int_{\partial M} L_{Bndy \
grav}$.  This simplest case occurs when $L_{Bndy \ grav}$ contains
no background structures. The allowed bulk diffeomorphisms which
leave the action invariant include diffeomorphisms deep in the bulk,
boundary diffeomorphisms, and, if $d$ is odd and $L_{Bndy \ grav}$
is conformally invariant, local Weyl transformations on the
boundary.  As for the Neumann case, we assume that $\partial \Sigma
:= \Sigma \cap \partial M$ is a compact manifold without boundary.
Since the dual theory is a gravity theory with compact spatial
slices, we expect the above transformations to again be gauge and
the associated charges to vanish.

It is, however, also interesting to allow $L_{Bndy \ grav}$ to
depend on a background structure, such as a fixed source $T_{ij}^{(S)}$.
Note that only transformations that leave the background structure
invariant are allowed. We consider the case where
\begin{equation}
S_{Bndy \ grav}  = \int_{\partial \cM} L_{diff \ inv}[g_{(0)}] +
\frac{1}{2}\int_{\partial \cM} \sqrt{g_{(0)}}\ T^{(S)}_{ij}g^{(0)ij},
\label{boundaryL}
\end{equation}
and where $L_{diff \ inv}[g_{(0)}]$ contains no background
structures. In this case our boundary condition becomes
\begin{equation}
0= T^{(total)ij} = T^{ij}  + \frac{2}{\sqrt{-g_{(0)}}}\varQ{L_{diff
\ inv}[g_{(0)}]}{g_{(0)ij}} - T_{(S)}^{ij}+\frac{1}{2}g^{(0)ij}
T_{\,\,k}^{(S)k}.\label{totalstress}
\end{equation}
This condition implies the useful result
\begin{eqnarray}
\label{eq:variationLB} \delta L_{Bndy \ grav} &=& \frac{\delta
L_{Bndy \ grav}}{\delta g_{(0)ij}} \delta g_{(0)ij} d^d x + \dH
\Theta_{Bndy \ grav} \cr &=& - \frac{\sqrt{-g_{(0)}}}{2} T^{ij}
\delta g_{(0)ij} d^d x + \dH \Theta_{Bndy \ grav}.
\end{eqnarray}

The infinitesimal charges are defined as before, but contain a term
from $\omega_{Bndy \ grav}$:
\begin{equation}
k_\xi \equiv -I_\xi \omega_{EH}[\cL_\xi g,\delta g]
+\omega_{ct}[\cL_\xi g,\delta g]+\omega_{Bndy \ grav}[\cL_\xi
g_{(0)},\delta g_{(0)}]\label{eq:12}.
\end{equation}
We have seen that the first two terms reduce on-shell to
\eqref{chargeLB}.  Since the source term in (\ref{boundaryL})
contains no derivatives, it does not contribute to the symplectic
structure and we may choose $\Theta_{Bndy \ grav}$ to be covariant
under diffeomorphisms which preserve the action \cite{Iyer:1994ys}.

Let us first discuss boundary diffeomorphisms. From $\dH i_\xi L_B =
\cL_\xi L_{diff \ inv} = \cL_\xi g_{(0)ij}$ $ \varQ{L_{diff \
inv}}{g_{(0)ij}}d^d x+\dH\Theta_{Bndy \ grav} [\cL_\xi g_{(0)}]$ and
\eqref{totalstress}, we observe that the Noether current $J_{\xi}
\equiv i_\xi L_{diff \ inv} - \Theta_{Bndy \ grav}[\cL_\xi g_{(0)}]$
satisfies
\begin{equation}
\dH J_{\xi} = -\frac{1}{2} \sqrt{-g_{(0)}}(T^{ij} -T_{(S)}^{ij}+\frac 1
2 T^{(S)i}_{\; i}g^{(0)ij}) \cL_\xi g_{(0)ij} d^d x.
\end{equation}
Applying the contracting homotopy $I_\xi$ and integrating over the
sphere at infinity yields
\begin{equation}
\oint_{\partial \Sigma} J_{\xi} = -\oint_{\partial \Sigma} \left(
\sqrt{-g_{(0)}}  (T_{ij} -T^{(S)}_{ij}+\frac 1 2 T^{(S)i}_{\;
i}g_{(0)ij})n^i[ g_{(0)}] \xi^j \right). \label{eq:JB}
\end{equation}
Now, the boundary symplectic structure satisfies
\begin{eqnarray}
\oint_{\partial \Sigma} \omega_{Bndy \ grav}[\cL_\xi g_{(0)},\delta g_{(0)}] &=& \oint_{\partial \Sigma}
\delta_{\cL_\xi g_{(0)}}\Theta_{Bndy \ grav} - \delta \Theta_{Bndy \ grav}(\cL_\xi g_{(0)})\\
&=& \oint_{\partial \Sigma} i_\xi \dH \Theta_{Bndy \ grav} - \delta \Theta_{Bndy \ grav}(\cL_\xi g_{(0)})\\
&& \hspace{-120pt} =\delta \oint_{\partial \Sigma} J_{\xi} +
\oint_{\partial \Sigma} i_\xi \left(\frac{1}{2}\sqrt{-g_{(0)}}
(T^{ij}-T_{(S)}^{ij}+\frac 1 2 T^{(S)i}_{\; i}g^{(0)ij} )\delta g_{(0)ij}
d^dx\right),
 \label{eq:omegaB}
\end{eqnarray}
where we have used the definition of $\Theta_{Bndy \ grav}$.
Inserting~\eqref{chargeLB}, \eqref{eq:JB} and \eqref{eq:omegaB} into
\eqref{eq:12}, we finally obtain
 \begin{eqnarray}
 \oint_{\partial \Sigma} k_\xi \approx \delta \oint_{\partial \Sigma}
 \left( \sqrt{-g_{(0)}} T^{(S)}_{ij} n^i[ g_{(0)}] \xi^j \right),
  \label{eq:chargesource}
 \end{eqnarray}
since the two first terms in \eqref{eq:omegaB} cancel those
in~\eqref{chargeLB}. The charges thus vanishes identically when
there is no source, and in this case all boundary diffeomorphisms
are gauge as expected. In the presence of a source, it is
straightforward to check that these charges are conserved.  This
follows from the fact that only diffeomorphisms satisfying $\cL_\xi
T^{(S)}_{ij} = 0$ are allowed and from the conservation equation $D^j
T^{(S)}_{ij} = \half D_i T^S$ which results from taking the divergence
of (\ref{totalstress}).  Some charges may now be non-zero. However,
for any $\xi$ with the property that $f\xi$ is also allowed for any
scalar function $f$ on the boundary, we may take $f=1$ on $\partial
\Sigma$ and $f=0$ on some other cut of the boundary. Conservation
then implies that the charge for $\xi$ vanishes and that $\xi$
generates a pure gauge transformation.

When $L_{Bndy \ grav}$ is Weyl invariant and $d$ is odd, one can
also consider boundary Weyl transformations; i.e., transformations
generated by \eqref{eq:conf} with arbitrary $\delta \sigma$ but
$\phi^i = 0$.    Let us call such a bulk vector field $\xi_W$, and
define $\delta_W g_{(0)ij}=2\delta \sigma g_{(0)ij}$.  Note that
Weyl invariance requires $T_{(S)}^{ij}g_{(0)ij}=0$ for all
$g_{(0)ij}$, so that in fact  $T^{(S)}_{ij} =0$.

Since both $T^{ij} g_{(0)ij}$ and $\delta_W L_{Bndy \  grav}$ vanish
in this context, $\dH\Theta_{Bndy \ grav} [\delta_W g_{(0)}] = 0$ by
 \eqref{eq:variationLB}. In addition, the Weyl variation of \eqref{eq:variationLB} implies
$0=-\delta_{W}(\frac{\sqrt{g_{(0)}}}{2}T^{ij}\delta
g_{(0)ij})+\dH\delta_W \Theta_B[\delta g_{(0)}]$. Since the first
term is zero by the transformation properties
\eqref{eq:34}-\eqref{eq:35}, $\delta_W \Theta_B[\delta g_{(0)}]$ is
also closed. Thus $\omega_B[\delta_W g_{(0)},\delta g_{(0)}]$ is
conserved for arbitrary $\delta \sigma$. Using the freedom to take
$\delta \sigma =0$ in the past, we conclude that
\begin{eqnarray}
\oint_{S^\infty} \omega_B[\delta_W g_{(0)},\delta g_{(0)}] &\approx&
0.
\end{eqnarray}
Moreover, because $\xi_W^i$ vanishes at the boundary (see
\eqref{eq:conf}) as $O(r^2)$,
 the contribution from \eqref{chargeLB} also
vanishes.  Thus the total charge for boundary Weyl transformations
vanishes identically on-shell.   We conclude that the expected
boundary diffeomorphisms and Weyl transformations are pure gauge.

\section{The Dynamics of boundary gravity}
\label{dynamics}

We have argued that string/M-theory with AdS${}_{d+1}$ Neumann or
mixed boundary conditions for the graviton is dual to a
$d$-dimensional CFT coupled to gravity.  It is natural to ask what
this correspondence can say about the dynamics of the boundary
gravity theory.  In particular one might like to know if the theory
is stable and/or UV-complete.

At least for odd $d$, it is natural to restrict discussion to the
Weyl-invariant Neumann theory (induced gravity only) and to its
relevant deformations by diffeomorphism-invariant operators.  While
we do not explore graviton loop effects here, it seems reasonable to
suppose that such theories are well-defined in the UV (though
they may have instabilities associated with ghosts or tachyons).
Due to its coupling to the stress tensor, the graviton has conformal
dimension zero for all $d$.  This can also be seen from the fact
that the graviton transforms as a tensor (with no extra conformal
weight) under the action of any conformal Killing field. As a
result, the relevant deformations are those containing less than or
equal to $d$ derivatives.  We shall also restrict discussion to such
terms for even $d$.

As usual, the simplest case arises for $d=2$.  There, the gravity
theory induced from any CFT is given by the non-local Polyakov
action, as can be uniquely determined by integrating the trace
anomaly \cite{Polyakov:1987zb,Polyakov:1989dm}.  Replacing the
original metric $g_{(0)ij}$ with $e^{2\sigma} \bar g_{(0)ij}$
introduces a new Weyl invariance and allows one to write the full
theory as the original CFT action evaluated on the Weyl-transformed
fields coupled to Liouville gravity.  In this description, the
central charge of the Liouville theory is of course equal in
magnitude but opposite in sign to that of the CFT so that the total
central charge vanishes. It is straightforward
\cite{Banados:1998gg,Skenderis:1999nb,Papadimitriou:2007sj} to
repeat this argument for an AdS${}_3$ bulk using the trace anomaly
of the boundary stress tensor $T_{ij}$. The fact the central charge
of the Liouville theory cancels that of the original CFT is then
apparent through the Neumann boundary condition (which requires
$T_{ij}=0$ so that the total conformal anomaly
vanishes)\footnote{There are also many other interesting connections
between AdS${}_3$ gravity and Liouville theory
\cite{Coussaert:1995zp,Bautier:1999ic,Bautier:2000mz,Carlip:2005tz},
typically of a form which relate the Liouville action to what for us
is the effective dynamics of the stress tensor as opposed to the
conformal mode of the boundary graviton. This Liouville field is
zero in the Neumann theory.}. Our results show that this 2d theory
is described by the bulk dynamics with Neumann boundary conditions.
Furthermore, the only diffeomorphism-invariant deformations with two
derivatives or less are boundary cosmological and Einstein terms,
the latter being purely topological and the former merely adding a
potential for the Liouville field.  The result is a well-defined
CFT, which is thus UV-complete.

The higher-dimensional case is more complicated.
Diffeomorphism-invariance no longer restricts local degrees of
freedom in the boundary metric to the conformal factor, and the
trace anomaly no longer determines a unique action for the induced
gravity theory, see e.g.
\cite{Deser:1976yx,Fradkin:1983tg,Riegert:1984kt,Antoniadis:1991fa,Deser:1993yx,Deser:2000un,Mazur:2001aa}.
An explicit calculation is in order. Below, we explore the dynamics
of the Neumann theory for all odd $d$ and for $d=4$ by calculating
the graviton propagator for perturbations about flat space. (See
also \cite{Skenderis:1999nb} for an attempt to study the effective
action for $g_{(0)}$ by calculating the bulk AdS action when
$g_{(0)}$ is conformally flat.)  For odd $d$, the theory is
both ghost- and tachyon-free. However, for $d=4$, there is a tachyon
with a ghost-like polarization.  Our effective action for $d=4$ is
consistent with previous results \cite{Liu:1998bu}.

For $d=3$ we also consider adding a gravitational Chern-Simons term
(the kinetic term of $d=3$ conformal gravity) on the boundary and a
boundary Einstein term.  The theory is perturbatively stable when
the boundary Newton's constant ($G_B$) is infinite or negative, but develops a ghost (and a
tachyon) for $G_B
> 0$. In this respect the $d=3$ Neumann theory is similar both to
topologically massive gravity
\cite{Deser:1981wh,Deser:1982vy,Deser:1982sv,Li:2008dq,Carlip:2008jk}.
In the limit where the bulk Newton's constant diverges, our system
becomes precisely TMG.

\subsection{The Neumann theory at the linear level}

For any dimension $d$, empty AdS${}_{d+1}$ expressed in Poincar\'e
coordinates has $T_{ij}=0$ (see e.g.,
\cite{Henningson:1998gx,Balasubramanian:1999re}).  As a result, it
provides a solution satisfying Neumann boundary conditions with the
flat boundary metric $\eta_{ij}$.  Let us denote this bulk metric by
$\bar g_{\mu \nu}$ and consider the perturbative expansion
$g_{\mu\nu} = \bar g_{\mu\nu} +h_{\mu\nu}$, using the
Fefferman-Graham-like gauge $h_{\mu 0} = 0$.

The details of the computations are relegated to the appendices.
First, in appendix \ref{app:action} we compute  the $d \ge 3$
effective action for $g_{(0)}$ and $T^{ij}$ (in the presence of a
source $T_{(S)}^{ij}$) at quadratic order in the perturbation. Our
calculations follow largely \cite{Liu:1998bu}, though we include all
counterterms.

The linearized stress-tensor (see \cite{deHaro:2000xn}) satisfies
\begin{eqnarray}
T_{ij} \approx \frac{d}{16\pi G}h_{(d)ij},\;\; \text{for $d$ odd,
and } \quad  T_{ij} \approx \frac{1}{8\pi G}(2 h_{(4)ij}+(3+2\alpha)
\tilde h_{(4)ij}),\;\; \text{for $d=4$.} \ \ \ \label{eq:Tij4}
\end{eqnarray}
Here, we have generalized the Neumann action (\ref{actionS}) to
include the additional boundary term $ \frac{\alpha}{16\pi G} \int
d^d x A_{(d)} $ which would naturally arise by changing the scale in
the $\log \epsilon$ counterterm via $\log \epsilon \rightarrow \log
\epsilon - \alpha$. Even in even dimensions,  the trace of
\eqref{eq:Tij4} vanishes as the anomaly is at least quadratic in the
curvature. The boundary condition (\ref{totalstress}) thus implies
\begin{equation}
T_{ij} \approx T^{(S)}_{ij}, \label{eq:boundaryEOM}
\end{equation}
but relating $T_{ij}$ to $g_{(0)ij}$ requires solving the bulk equations of motion.
The solutions are discussed in appendix
\ref{lingrav} using a Fourier expansion $h_{ij}(x_0,x_k) = \int \frac{d^d k}{(2\pi)^d} e^{i \eta^{ij} k_j x_i} \hat h_{ij}(x_0,k^i)$, and similarly for $T^{(S)}_{ij}$.

The two-point function then follows by taking the derivative of this
on-shell action with respect to the source as\footnote{The overall
plus sign follows from our convention for the sign of the action,
and the factor $4$ comes from our $1/2$ coupling \eqref{boundaryL}
of the source to $g_{(0)}$.}
\begin{equation}
\langle h_{(0)ij}(x^m)\, h_{(0)kl}(x^n) \rangle \equiv +4 \left(\Q{}{T^{ij}_{(S)}}\Q{}{T^{kl}_{(S)}} S_{ren}^{(S)}\right)_{T_{(S)}^{ij}=0}.
\label{defhh}
\end{equation}
Because the linearized stress-tensor $T^{ij}$ is transverse and
traceless, the propagator can be expressed in terms of the tensor
\begin{eqnarray}
\Pi^{(2)}_{ij,kl}(k)&=&\frac12\left[ \Pi_{ik}(k) \Pi_{jl}(k)+ \Pi_{il}(k) \Pi_{jk}(k)-\frac{2}{d-1} \Pi_{ij}(k) \Pi_{kl}(k)\right]\label{tensorPi2}
\end{eqnarray}
where $\Pi_{ij}(k) = \eta_{ij} - \frac{k_i k_j}{k^2}$.

Plugging the solutions \eqref{eq:oddPsi}-\eqref{eq:evenpsiE} into
the on-shell action \eqref{eq:SlinS}, we obtain from \eqref{defhh}
the Euclidean Neumann propagators,
\begin{equation}
 \langle \hat h_{(0)ij}(k) \hat h_{(0)kl}(-k) \rangle \approx
\frac{2^{d+4} (\Gamma(\frac d 2))^2\, G}{ (-1)^{\frac{d-1}{2}} }
\frac{\Pi^{(2)}_{ij,kl}(k)}{(k^2)^{\frac d 2}}.\label{propd3}
\end{equation}
for all odd $d$ and
\begin{equation}
 \langle \hat h_{(0)ij}(k) \hat h_{(0)kl}(-k) \rangle \approx
256\pi G  \frac{\Pi^{(2)}_{ij,kl}(k)}{k^4 \log (k^2 L^2)}\label{ex:prop}
\end{equation}
with $L \equiv  \frac{l}{2}e^{\gamma+\frac \alpha 2}$ for $d = 4$.

We now make several comments on these results.   We observe that the
$d=4$ Euclidean propagator \eqref{ex:prop} has a pole at $k=1/L$
corresponding to a normalizable mode (see appendix \ref{lingrav}).
Thus the $d=4$ flat space boundary yields a tachyon whose mass
rescales with $\alpha$; i.e., with a change in the renormalization
scheme. Because this tachyon is transverse traceless, it corresponds
to both ghosts and normal tachyons depending on the polarization.
 Similar results also hold for even $d > 4$, see appendix
\ref{sec:ghosts}.  On the other hand, we show in appendix
\ref{sec:ghosts} that timelike propagating modes (i.e. with $m^2 =
\sqrt{-k^2} \geq 0$) are non-ghost.  For odd $d$, all propagating
Neumann modes have $m^2
> 0$ and these theories are ghost-free. Instead of analyzing the
Euclidean propagator, we find it convenient to demonstrate these
results in Lorentz signature by computing the symplectic norm of
positive frequency modes.

It is clear that the effective action is in general a non-local
functional of the boundary metric. In odd dimensions, the Lagrangian
has the form
\begin{equation}
S^{d\;\text{odd}}_{Neu}[g_{(0)}] \approx S^{d\;\text{odd}}_{Neu}[\eta] + \frac{(-1)^{\frac{d-1}{2}}}{2^{d+5} (\Gamma(\frac{d}{2}))^2 G} \int d^d x  h^{TT}_{(0)ij} (\d^2)^{\frac{d}{2}} h^{ij}_{(0)TT} +O(h^3_{(0)}) .\label{eq:effactionodd_d}
\end{equation}
It would be interesting to compare the normalization of the
quadratic term with computations for the dual strongly coupled gauge
theory.

For $d=4$, the logarithmic behavior of \eqref{ex:prop} is familiar
from early AdS/CFT results, see \cite{Gubser:1998bc} and
\cite{Hawking:2000kj,Hawking:2000bb}, and from studies of
trace-anomaly contributions to $d=4$ gravity
\cite{Tomboulis:1977jk}. Our propagator merely reproduces the
graviton part of the strong coupling, large $N$ check obtained in
\cite{Liu:1998bu} of the claim that bulk $\cN = 8$ gauged
supergravity leads to a boundary effective action for $\cN =
4$ conformal supergravity fields that agrees with the effective
action for $\cN = 4$ super-Yang-Mills in conformal supergravity
backgrounds.  In particular,
\begin{equation}
S^{d=4}_{Neu}[g_{(0)}] \approx S^{d=4}_{Neu}[\eta] +
\frac{1}{2^5\,16\pi G}\int d^4 x \d^2 h^{TT}_{(0)ij} \log(-L^2 \d^2)
\d^2 h^{ij}_{(0)TT} +O(h^3_{(0)}). \label{eq:effaction4}
\end{equation}
Matching the normalization factor in \eqref{eq:effaction4} to the
super-Yang-Mills result is most easily done using the relation
$\frac{l^3}{16 \pi G} = \frac{N^2}{8\pi^2}$ \cite{Henningson:1998gx}
obtained from the conformal anomalies.  We also note that
\eqref{eq:effaction4} agrees at quadratic order with the non-local
effective action for the anomaly in $d=4$ advocated in
\cite{Deser:1976yx,Deser:1993yx,Liu:1998bu,Deser:2000un}.

\subsection{$d=3$ and topologically massive gravity}

Our discussion above focused on the pure Neumann theory; i.e., the
case where the $d$-dimensional gravitational dynamics is just that
induced by the CFT and where no explicit gravitational terms have
been added to the action.  We have computed the propagator to
leading order in the $1/N$ approximation, where we have seen for odd
$d$ that the theory contains no ghosts or tachyons; Minkowski space
appears to be a stable solution of the theory.  Also for odd $d$,
Weyl-invariance encourages the belief that the theory is
well-defined in the UV. It is therefore interesting to ask about
relevant deformations of this theory.

We will investigate the case $d=3$, where there are three
interesting deformations. They are the (marginal) gravitational
Chern-Simons term, the (relevant) Einstein-Hilbert term, and the
(relevant) cosmological constant term.  By themselves, this
collection of terms defines the theory of (cosmological)
topologically massive gravity \cite{Deser:1981wh,Deser:1982vy}.  We
will confine ourselves to the case where the boundary cosmological
constant vanishes so that $AdS_4$ in Poincar\'e coordinates remains
a valid solution of the theory.  The total (Euclidean) action that
we consider is therefore the bulk action \eqref{actionS} supplemented by the boundary action \eqref{boundaryL} with dynamics
\begin{eqnarray}
\int_{\partial \cM} L_{diff \ inv}[g_{(0)}]& =& \frac{\lambda\, l^2}{16\pi G}\int_\cM d^4 x\frac{1}{8}\eps^{\mu\nu\alpha\beta}R_{\mu\nu\rho\sigma}R_{\alpha \beta}^{\;\;\;\;\rho\sigma} + \frac{1}{16\pi G_B}\int_{\d\cM} d^3 x
\sqrt{-g_{(0)}}R_{(0)}, \ \ \  \label{actiontotCG}
\end{eqnarray}
where we have written the boundary gravitational Chern-Simons term
as a bulk integral over the Pontryagin density with $\lambda$ dimensionless and we have chosen to write the Einstein term with
the same sign convention as for the bulk Einstein term.

Setting $l=1$, the boundary condition \eqref{totalstress} is
given by
\begin{equation}
\label{Ctotstress}
\frac{3}{16\pi G} g_{(3)ij} -
\frac{\lambda}{8\pi G} C_{(0)ij} - \frac{1}{8\pi G_B}  G_{(0)ij}
\approx T_{(S)ij}-\half g_{(0)ij} g_{(0)}^{kl} T_{(S)kl},
\end{equation}
where $C_{(0)}^{ij} =
\frac{1}{\sqrt{g_{(0)}}}\epsilon^{ikl}D_{k}(R^j_{(0)\;l}- \frac 1 4
\delta^j_l R_{(0)})$ is the Cotton tensor and we have explicitly
evaluated the stress-tensor $T_{ij} = \frac{3}{16\pi G}g_{(3)ij}$.

As a brief aside, consider the limit $G_B \rightarrow \infty$ (which implies $T_{(S)i}^i \rightarrow 0$) in which Weyl invariance is restored.   We note that the Weyl tensor of the bulk metric expanded in Fefferman-Graham coordinates \eqref{eq:FG2} is determined by the leading behavior of its electric and magnetic parts,
\begin{eqnarray}
\cE_{ij} \equiv r C_{r i  j r}|_{r=0} =  8\pi G\ T_{ij}, \qquad \cB_{ij} \equiv  \frac{r}{2} \epsilon_{i}^{\;\,kl} C_{r jkl}|_{r=0} = C_{(0)ij},
\end{eqnarray}
Both $\cE_{ij}$ and $\cB_{ij}$ are symmetric, traceless and covariantly conserved with respect to $g_{(0)}$. Therefore, Neumann boundary conditions (for $\lambda = 0$) are equivalent to fixing the electric part of the Weyl tensor, while Dirichlet boundary conditions, recovered in the limit $\lambda \rightarrow \infty$, are equivalent to fixing the magnetic part of the Weyl tensor.  Thus, switching from Neumann to Dirichlet boundary conditions corresponds to a sort of gravitational electric-magnetic duality \cite{Henneaux:2004jw,Julia:2005ze,Leigh:2007wf,deHaro:2007fg} and choosing a more general $\lambda$ might be though of as a duality rotation.  In addition,
in Euclidean signature and when $\lambda = \pm 1$, the boundary conditions are equivalent to the (anti-)self-duality condition for the Weyl tensor (see also \cite{deHaro:2007fg}).

As described in the appendix \ref{lingrav}, the linearized solutions
around $AdS_4$ in transverse gauge can be decomposed into two types
of modes, transverse-trace and transverse-traceless. The
transverse-trace mode can be completely gauged away when the
boundary theory is Weyl invariant. When $G_B \neq 0$, Weyl
invariance is broken.  However, this mode still does not involve
$T_{ij}$ (since the trace anomaly vanishes for $d=3$), nor does it
depend on $\lambda$ (due to the Weyl invariance of conformal
gravity).  Using the constraint on the source $D^j T_{(S)ij}=\half
D_i T_{(S)}$, the transverse-trace part of the propagator is found
to be
\begin{equation}
 \langle \hat h_{(0)ij}(k) \hat h_{(0)kl}(-k) \rangle_{T,Tr\, part} \approx
-\frac{16\pi G_B}{k^2} \Pi_{ij}\Pi_{kl}\label{ex:propTrT}
\end{equation}
Despite the pole for $k^2 = 0$, there are no propagating modes in the Lorentz signature, because this mode is not a solution of the (linearized) boundary equations of motion \eqref{Ctotstress} when the source term is put to zero.


On the other hand, the transverse-traceless modes must satisfy the
linearized version of \eqref{Ctotstress} with transverse-traceless
$T^{(S)}_{ij}$. In Fourier space we have
\begin{equation}
k^2 \hat h_{(0)ij}^{TT}(k) + \frac{i\lambda G_B}{2G}(\epsilon_i^{\;\,kl} k_k k^2 \hat h^{TT}_{(0)lj}(k) + (i \leftrightarrow j)) = -16\pi G_B \hat T^{(S),TT}_{ij}(k) + \frac{3G_B}{G}\hat h_{(3)ij}(k).\label{eq:bndeqconfgrav}
\end{equation}
In Lorentzian signature (with vanishing source term), this boundary equation of motion admits a solution for any mass $m^2 \equiv -k^2 = k_1^2 -k_2^2-k_3^3$ that relates $\hat h_{(3)ij}(k)$ to  $\hat h_{(0)ij}^{TT}(k)$ (see the general bulk solution \eqref{generallinsol}).

For Euclidean signature one finds
\begin{eqnarray}
\label{413} \hat h_{(3)ij}(k) &=&
\frac{k^3}{3}\hat h_{(0)ij}^{TT},\\
\hat h_{(0)ij}^{TT}(k) &=& \frac{16\pi G}{k^2\left((k-\frac{G}{G_B })^2 - \vareps \lambda^2 k^2 \right) }
\left( (k-\frac{G}{G_B }) \hat T^{(S),TT}_{ij}(k)+ \frac{i \lambda}{2} ( \epsilon_i^{\;\,kl} k_k  \hat
T^{(S),TT}_{lj}(k) + (i \leftrightarrow j) \right) \nonumber
\end{eqnarray}
where the bulk solution is given by (\ref{eq:oddPsi}) and
$\vareps \equiv \eps_{ijk}\eps^{ijk}=+1$ in our Euclidean
conventions, though we will have $\vareps =-1$ in Lorentz
signature. The relation \eqref{413} reduces to the Neumann result
when \eqref{Tijodd} when $T_{(S)ij} \approx T_{ij}$ in the limit
$\lambda \rightarrow 0$, $G_B \rightarrow \infty$.

Computing \eqref{defhh} using the Neumann on-shell action (see
\eqref{eq:SlinS}) supplemented by the boundary dynamics in
(\ref{actiontotCG}) yields the part of the propagator involving
transverse-traceless modes,
\begin{eqnarray}
\label{3prop} \hspace{-30pt}\langle h_{(0)ij}(-k) h_{(0)kl}(k) \rangle_{TT\;part} &\approx& \frac{-32\pi G}{k^2\left((k-\frac{G}{G_B })^2 - \vareps \lambda^2 k^2 \right) } \left( (k-\frac{G}{G_B })   \Pi^{(2)}_{ij,kl} + \lambda\, k  \Pi^{(1.5)}_{ij,kl}\right),
\end{eqnarray}
where $\Pi^{(2)}_{ij,kl}$ is defined in \eqref{tensorPi2} and for
$d=3$ we follow \cite{Leigh:2003ez} in defining
\begin{eqnarray}
\Pi^{(1.5)}_{ij,kl}(k)&=&  \frac{i}{4}\left[\epsilon_{ikm}\Pi_{jl}(k)
+\epsilon_{jkm}\Pi_{il}(k) +\epsilon_{ilm}\Pi_{jk}(k)
+\epsilon_{jlm}\Pi_{ik}(k)\right]\frac{k^m}{k}\,.
\end{eqnarray}

There are now several interesting limits to consider. First, in the
limit $G \rightarrow \infty$ with $\frac{\lambda G_B}{G} =
\frac{1}{\mu}$ fixed, the bulk theory decouples and only the action
$S_{Bndy\ grav}$ remains.  This is precisely topologically massive
gravity (with $\Lambda =0$) and our propagator
\begin{eqnarray}
\frac{1}{16\pi G_B}\langle h_{(0)ij}(-k) h_{(0)kl}(k) \rangle
&\approx&  -\frac{1}{k^2} \Pi_{ij} \Pi_{kl} + \frac{2}{k^2(1-\vareps
\frac{k^2}{\mu^2}) } \left( \Pi^{(2)}_{ij,kl} - \frac{k}{\mu}
\Pi^{(1.5)}_{ij,kl}\right)
\end{eqnarray}
indeed agrees with \cite{Deser:1981wh,Deser:1982vy}. The massive
graviton of mass $|\mu|$ is a ghost for $G_B > 0$ and is a normal
mode for $G_B < 0$.

Second, we note that conformal invariance is restored in the limit
$G_B \rightarrow \infty$. The propagator then has a $1/k^3$ behavior
similar to that of the Neumann case \eqref{propd3} for $d=3$, but
the modes involve both transverse-traceless structures $\Pi^{(2)}$
and $\Pi^{(1.5)}$ with respective weight $\frac{1}{\vareps
\lambda^2-1}$ and $\frac{\lambda}{\vareps \lambda^2-1}$. In this
limit our graviton propagator can be thought of as an intermediate
step in the S-duality operation of \cite{Leigh:2003ez,Leigh:2007wf}.
For Euclidean signature ($\vareps = +1$), there is a chiral part of
$h^{TT}_{(0)ij}$ that becomes pure gauge when $\lambda = \pm 1$, as
may be seen by the fact that it solves the boundary equations of
motion with vanishing source for any $k^i$. The resulting linearized solutions describe
regular instantons as remarked in \cite{deHaro:2007fg}. In Lorentz
signature the additional gauge happens at $\lambda = \pm i$ and there is no special
real value of $\lambda$.

Finally, consider the low momentum limit $k \ll \frac{G}{G_B l}$,
 $k \ll \frac{G}{\lambda G_B l}$ in which only the $d=3$ massless
modes remain. Here the propagator reduces to that of $d=3$ Einstein
gravity.  This is just the expected renormalization-group flow:  the
Einstein term is a relevant deformation of conformal gravity and
dominates the dynamics in the infrared independent of the UV
dynamics.

Let us now consider the propagator (\ref{3prop}) in more detail.
There are two poles at $k = \frac{G}{G_B(1\pm \sqrt{\vareps
\lambda^2})}$ that coincide when $\lambda = 0$. A careful check of
the linearized solutions confirms that in order to have regular
solutions in the interior (when $x_0 \rightarrow \infty)$, our $k$
must have non-negative real part. There is no physical excitation
corresponding to poles with $\text{Re } k < 0$.  There is therefore
no Lorentzian solution with $k^2 > 0$, i.e. no tachyon, when $G /
G_B <0$.

As in the Neumann theory, we can compute the ghost spectrum by
analyzing the symplectic norm of Lorentzian positive frequency modes
satisfying the boundary conditions \eqref{eq:bndeqconfgrav}. The
mode analyzed in \cite{Deser:1981wh,Deser:1982vy} is a ghost for
$G_B
> 0$ but is ghost-free for $G_B <0$. Using the tools of the appendix
\ref{sec:ghosts}, it is straightforward to show that all other
timelike modes are free of ghosts. Thus for Lorentz signature
$\vareps = -1$ and the usual choice $G > 0$, the theory is both
ghost- and tachyon-free and is perturbatively stable for $G_B < 0$.

\section{Discussion}

\label{disc}

In the context of asymptotically AdS${}_{d+1}$ gravity for
$d=2,3,\text{ or }4$, we have shown that counter-term contributions
to the bulk symplectic structure render normalizable all
fluctuations of the leading Fefferman-Graham coefficient
$g_{(0)ij}$. The same result is expected to hold for all $d \ge 2$.
This allows one to consider a variety of new boundary conditions for
AdS gravity. Under the AdS/CFT correspondence, our Neumann boundary
conditions are dual to the induced gravity theory associated with
the dual CFT, while other boundary conditions are dual to coupling
this CFT theories with explicit gravitational terms.  In particular,
for $d=2$ the effective action for $g_{(0)ij}$ is that of Liouville
gravity \cite{Banados:1998gg,Skenderis:1999nb,Papadimitriou:2007sj}.
We now see that, writing $g_{(0)ij} = e^{2\sigma} \bar g_{(0)ij} $,
there is in fact a AdS/CFT-like duality between AdS${}_3$ with
Neumann boundary conditions and the appropriate CFT on $\bar
g_{(0)ij}$ coupled to a Liouville theory for $\sigma$. We expect our
methods to be similarly useful in interpreting the calculations of
\cite{Freivogel:2006xu}.

 We also briefly analyzed the dynamics associated with our
boundary conditions by computing boundary graviton two-point
functions. When expanded about empty AdS${}_{d+1}$, the Neumann
theories for odd $d$ are both ghost- and tachyon-free, though the
theories for even $d \ge 4$ contain both tachyons and tachyonic
ghosts. For $d=4$ our results follow from those of
\cite{Gubser:1998bc,Liu:1998bu,Hawking:2000bb}. For $d=3$ we
considered adding boundary Einstein and Chern-Simons terms.  This
theory is much like topologically massive gravity (TMG), even when
the Chern-Simons coupling vanishes. For $G_B/G > 0$ the theory is
perturbatively unstable.  It is perturbatively stable for $G >0$ and
$G_B < 0$, though in the presence of a boundary cosmological
constant the theory will contain negative energy BTZ boundary black
holes (again like TMG \cite{Li:2008dq}).

It is interesting to reconsider the results of
\cite{Ishibashi:2004wx} in light of the above understanding of
symplectic structures. The authors of \cite{Ishibashi:2004wx}
analyzed the normalizability of AdS$_{d+1}$ gravitational modes for
all $d$ using an inner product that arose naturally in their study
of the equations of motion. They found that general fluctuations
were normalizable at infinity for $d=3$, but not for other values of
$d$. In retrospect it is clear that their inner product did not
in fact correspond to the usual symplectic structure $\omega_{EH}$
of Einstein-Hilbert gravity, but instead differed from $\omega_{EH}$
by a boundary term that is closely related to our $\omega_{ct}$ for
$d=3$ but not for $d \ge 4$. Because the gauge-invariant variables
used in their work become trivial for $d=2$, they found no new
normalizable modes for this case.

The lesson is then that lack of normalizability of certain bulk
modes with respect to a given symplectic structure need not rule out
construction of a theory in which such fluctuations are allowed.
With this new perspective, one would like to again investigate the
correlation found in \cite{Klebanov:1999tb} between scalar field
modes whose usual Klein-Gordon norm diverges and those which would
correspond to an operator of dimension small enough to violate the
CFT unitarity bound.  As in our case, the  action for tachyonic
scalars can be made finite for such slow fall-off modes by adding
counter-terms containing derivatives, which will in turn contribute
to the symplectic structure.  For such cases the CFT unitarity bound
predicts that this sort of renormalization for slow fall-off scalar
modes always lead to ghosts.

Many other directions also merit exploration. First, supersymmetric
extensions in the AdS${}_4 \times S^7$ theory should be
straightforward and ghost-free. For this theory, all scalars in the 4-dimensional
graviton supermultiplet  have masses  in the Breitenlohner-Freedman
range \cite{AdS4S7}; see \cite{Breitenlohner:1982bm,Hollands:2006zu}
for discussions of multi-trace boundary conditions for the $\cN=1$
scalar super-multiplet.  Second, though our general methods should apply, one expects novel
features to arise in the case where the field theory lives on a
spacetime with interesting boundaries. With a negative boundary
cosmological constant, such settings should lead to multi-layered
AdS/CFT-type dualities. One might also investigate further
gravitational consequences of the trace anomaly, e.g. such as those
suggested in \cite{Mazur:2001aa,Mottola:2006ew}, using bulk
techniques (see \cite{Hawking:2000kj,Hawking:2000bb}, naturally
interpreted in our framework without the need for a brane).

Most importantly, however, one would like to use bulk techniques to
gain a better understanding of the ghosts that arise in the boundary
theories. One possibility is that these ghosts condense in an
interesting way, or that strong-coupling effects save the day
\cite{Kaku:1982xt,Tomboulis:1983sw}.  If this occurs for some
boundary condition, one would expect the associated boundary theory
to be UV-complete. Another possibility, however, is that the
boundary theory admits a UV-completion via some (ghost-free) string
theory. The full boundary string theory might then be dual to the
original asymptotically AdS string theory when appropriate boundary
conditions are imposed on the full set of massive string modes. Such
a correspondence could provide a substantial enlargement of the
known set of string dualities.

\subsection*{Acknowledgements}
The authors would like to thank Maria Jose Rodriguez, who worked
with them on some early stages of the project. DM also thanks
Stefan Hollands and Simon Ross for many discussions of related
issues and Murat G\"unaydin for useful correspondence. GC thanks S. Hartnoll, M. Headrick, D. Mateos, E. Mottola, R. Porto and M.
Romo, E. Silverstein and L. Susskind for their comments and
K. Skenderis, P. Spindel and S. Deser for their correspondence.
This work was supported in part by the US National Science
Foundation under Grant No.~PHY05-55669, and by funds from the
University of California. GC was supported as David and Alice van
Buuren Fellow of the BAEF foundation.

\appendix

\section{The Symplectic form of $d=4$ Weyl gravity}
\label{app:Weylomega}

This appendix computes the symplectic structure of $d=4$
Weyl-squared gravity and finds that it can be expressed in terms of
the Einstein-Hilbert symplectic structure. This calculation is
central to our study of $d=4$ as the anomaly term \eqref{eq:anomal}
is the sum of $\frac{1}{16}\sqrt{g_{(0)}}E_{(4)}$, the Euler term
for $g_{(0)}$ whose variation is a total derivative, and the
Lagrangian density for Weyl-squared gravity in four dimensions,
\begin{equation}
L^{Weyl} = -\frac{1}{16} \sqrt{-g} C_{ijkl} C_{}^{ijkl},\label{eq:LagrWeyl}
\end{equation}
evaluated on the metric $g_{(0)}$. Here the Weyl tensor is
$C^i_{jkl} =
R^i_{jkl}-2(\delta^i_{[k}K_{l]j}-g_{j[k}K_{l]}^{\;\,i})$ in terms of
the tensor
\begin{equation}
K_{ij}= \frac{1}{2}(R_{ij}-\frac{1}{6}g_{ij}R).\label{Kab}
\end{equation}
The symplectic structure $\omega_{A_{(4)}}[\delta g_{(0)},\delta
g_{(0)}]$ is thus identical to that of Weyl gravity
\eqref{eq:LagrWeyl}, since the Euler term will not contribute.

Let us first compute the boundary term coming from the variation of
the anomaly, $\delta A_{(4)} = (\cdots)\,\delta g_{(0)} + \d_k
\Theta^k_{A_{(4)}}[\delta g_{(0)}]$, where $(\cdots)$ denotes the
equations of motion of Weyl gravity. We have
\begin{eqnarray}
\Theta^k_{A_{(4)}}[\delta g] &=& \frac{\sqrt{-g}}{4}\Big( (-2K^{ij}+g^{ij}K)\delta \Gamma^k_{ij} + (2K^{kj}-g^{kj}K)\delta \Gamma^i_{ij} \nonumber
\\ &&\hspace{-35pt}+(2 D^j K^{ik}-D^k K^{ij}-g^{ik}D^j K -g^{ij} ( D_l K^{kl} -D^k K) )\delta g_{ij}  \Big). \label{BigThet}
\end{eqnarray}
In place of computing $\omega^k_{Weyl} \equiv \delta
\Theta^k_{A_{(4)}}[\delta g]$, we find it more convenient to compute
 the second order term in the expansion of the bulk
symplectic structure \eqref{eq:omegaEHgamma} and then link the
result to $\omega^k_{Weyl}$.   Note that the tensor
$P^{abcdef}[\gamma]$ \eqref{tensorP} admits a Fefferman-Graham
expansion of the form $P^{abcdef}[\gamma] = x_0^{-d+6}
P^{abcdef}[g_{(0)}]+x_0^{-d+8}
P_{(2)}^{abcdef}[g_{(0)},g_{(2)}]+O(x_0^{-d+9})$ where the second
term is given by
\begin{eqnarray}
P^{abcdef}_{(2)}[g_{(0)},g_{(2)}] &= &\half g_{(2)i}^i P^{abcdef}_{(0)} + \half \Big[ G_{(0)}^{abc(e}g_{(2)}^{f)d}-\half G_{(0)}^{abcd}g^{ef}_{(2)} + (c \leftrightarrow d) \nonumber \\
&&+ ((ab) \rightarrow (cd) \rightarrow (ef) \rightarrow (ab)) \Big],
\end{eqnarray}
in which one includes all terms generated by cyclic permutations of
 the pairs of indices $(ab)$, $(cd)$ and $(ef)$, $G_{(0)}^{abcd} = \frac{\sqrt{-g_{(0)}}}{16\pi
G}(g^{a(c}g^{d)b}-g^{ab}g^{cd})$ and all indices are raised and
lowered with $g_{(0)}$.  As a result, the Einstein-Hilbert
symplectic form for $\gamma$ admits an expansion given in
\eqref{eq:omegaEHgamma} whose second order term is given by
\begin{eqnarray}
&&\omega^k_{(2)EH}[\delta g_{(0)},\delta g_{(0),(2)}] = -P^{kbcdef}_{(2)}\delta_2 g_{(0)cd}D_b \delta_1 g_{(0)ef} - P^{kbcdef}_{(0)}\Big( \label{omega4} \\
&& \delta_2 g_{(2)cd}D_b \delta_1 g_{(0)ef}+\delta_2 g_{(0)cd}D_b \delta_1 g_{(2)ef}-2 \delta_2 g_{(0)cd}\Gamma^i_{(2)be} \delta_1 g_{(0)fi} -(1 \leftrightarrow 2) \Big),\nonumber
\end{eqnarray}
where $\Gamma^a_{(2)bc} = D_{(b}g^a_{(2)\,c)}-\half D^a g_{(2)bc}$,
and the covariant derivative is that of $g_{(0)}$.

Now, it turns out that for $d=4$ we have $g_{(2)ab} = - K_{(0)ab}$
on-shell, where $K_{ab}$ is given by \eqref{Kab}. With these
formulae in hand, it is straightforward to show that \eqref{BigThet}
and \eqref{omega4} are related by
\begin{eqnarray}
\omega^k_{(2)EH}[\delta g_{(0)},\delta g_{(0),(2)}]|_{g_{(2)ab} = -K_{(0)ab}} &=& \frac{1}{8\pi G} (\delta_2 \Theta^k_{A_{(4)}}[\delta_1 g_{(0)}] - (1\leftrightarrow 2) ) \nonumber\\
&\equiv & -\frac{1}{8\pi G} \omega_{Weyl}[\delta g_{(0)},\delta g_{(0)}].
\end{eqnarray}
In computing this final result  we have discarded a boundary term of
the form $\d_l B^{[kl]}$ which is irrelevant for our purposes.

\section{The renormalized on-shell action}
\label{app:action}

This appendix computes the $d \ge 3$ effective action for $g_{(0)}$
and $T^{ij}$ at the boundary at quadratic order in a perturbative
expansion $g_{\mu \nu} = \bar g_{\mu \nu} + h_{\mu \nu}$ around
Poincar\'e AdS$_{d+1}$.  The background metric is
\begin{equation}
ds^2 = \bar g_{\mu\nu} dx^\mu dx^\nu = x_0^{-2} \eta_{\mu\nu}dx^\mu
dx^\nu, \qquad \mu,\nu=0,\,1,\,\dots,d,\label{eq:Poincare}
\end{equation}
and satisfies $\bar R_{\mu\nu} = -d \bar g_{\mu\nu}$, $\bar R =
-d(d+1)$.   Here  $\eta_{\mu\nu}dx^\mu dx^\nu= dx_0^2 + \dots +
dx_{d}^2$ in Euclidean signature, and $\eta_{\mu\nu}dx^\mu dx^\nu=
dx_0^2 - dx_1^2 + dx_2^2 + \dots + dx_{d}^2$ in Lorentzian
signature.  Below, we impose the Fefferman-Graham-like gauge $h_{\mu
0}=0$.

It is convenient to first expand the truncated action
\begin{equation}
S_{trunc} = \int_\cM  L_{EH} + \int_{\partial \cM} L_{GH} -
\frac{d-1}{8\pi G l} \int_{\partial \cM} d^d x
\sqrt{-\gamma},\label{Strunc}
\end{equation}
which includes only the first counterterm. One has
$\sqrt{g}(R-2\Lambda)=-2d\sqrt{\bar g}+\cL_2+\sqrt{\bar g}\bar
D_\alpha t^\alpha+O(h^3)$, where the linearized graviton Lagrangian
($\cL_2$) in AdS space  is
\begin{eqnarray}
\cL_2 &=& \frac{1}{2}\sqrt{\bar g}\big[ \frac{1}{2} \bar D_\mu h \bar D^\mu h - \bar D_\mu h \bar D^\nu h^\mu_{\;\nu}+\bar D_\mu h^{\alpha\beta}\bar D_\alpha h^\mu_{\;\beta} \nonumber\\ && \qquad - \frac 1 2 \bar D_\mu h_{\alpha\beta} \bar D^\mu h^{\alpha \beta}
 +d(\frac 1 2 h^2 - h^{\mu\nu}h_{\mu\nu})\big],\label{L2}
\end{eqnarray}
$h=\bar g^{\mu\nu} h_{\mu\nu}$ and, on-shell, $\cL_2 \approx
\sqrt{\bar g}D_\alpha v^\alpha$. The explicit expressions for
$t^\alpha$ and $v^\alpha$ are given in \cite{Liu:1998bu}. On-shell,
one can thus express $\int_\cM d^{d+1}x (\cL_2 + \sqrt{\bar g}\bar
D_\alpha t^\alpha)$ as a boundary term, and combine it with the
expansion of the boundary terms in \eqref{Strunc}. Using the gauge
condition $h_{0\mu}=0$, one finds \footnote{This result differs from
\cite{Liu:1998bu} by an additional trace term (and the term
proportional to $\d_j h^i_{\;0}$ in equation (4.16) of
\cite{Liu:1998bu} has been gauged fixed to zero). The rest of the
computation in \cite{Liu:1998bu} is independent on this additional
term since it vanishes in their subsequent analysis. In our context,
however, this additional term gives a divergent contribution that
cancels against contributions from other counterterms below.}
\begin{equation}
S_{trunc} \approx   - \frac{2 d}{16\pi G} \int_\cM d^{d+1}x
\sqrt{\bar g} +  \frac{1}{16\pi G} \int_{\partial \cM}d^d x
x_0^{1-d}\left( \frac{1}{4} h^j_{\;i}\d_0 h^i_{\;j}-\frac{1}{4}h
\d_0 h\right)  +O(h^3),\label{S2}
\end{equation}
where $h^i_{\;j} = x_0^2 \eta^{ik} h_{kj}$.

The Fefferman-Graham expansion for $h_{ij}$ is
\begin{eqnarray}
h_{ij} = x_0^{-2} h_{(0)ij} + h_{(2)ij}+\dots +x_0^{d-2} h_{(d)ij}+x_0^{d-2} \log x_0^2 \tilde h_{(d)ij}+ O(x_0^{d-1}),\label{eq:FGexph}
\end{eqnarray}
where the log term is present only for $d$ even (for $d=2$, $ \tilde
h_{(d)jk} =0$ \cite{deHaro:2000xn}). Using (\ref{eq:FGexph}), one
can rewrite the action \eqref{S2} as a finite piece plus a divergent
piece. In odd dimensions, the counterterms provide a minimal
subtraction of all divergent terms. The part of the renormalized
action \eqref{actionS} quadratic in the perturbation is then easily
obtained by identifying the finite term. For odd $d$ we find
\begin{eqnarray}
S^{quad}_{Neu} &\approx & \frac{1}{4} \int_{\partial \cM} d^d x
T_{ij} G^{ijkl} h_{(0)kl},\label{eq:Slin}
\end{eqnarray}
where $T_{ij}$ is the (linearized) stress-tensor $T_{ij} = \frac{d}{16\pi G}h_{(d)ij}$ where indices have been lowered with $\eta_{ij}$ and we
have defined for convenience the tensor $G^{ijkl} =
\eta^{i(k}\eta^{l)j} - \eta^{ij}\eta^{kl}$.

For even $d$, the counterterms contribute to the finite piece of the renormalized action. In four dimensions, this is manifest when the Einstein action for $\gamma$ is expanded as
\begin{equation}
\sqrt{-\gamma} R[\gamma] = r^{-2} \sqrt{-g_{(0)}}R_{(0)} +\frac{1}{2}\sqrt{-g_{(0)}}(R^{ij}_{(0)}R_{ij}^{(0)}-\frac 1 3 R^2_{(0)})+O(r^2).
\label{eq:43}
\end{equation}
We remark that the second order term is proportional to the anomaly
action \eqref{eq:anomal}. We will denote the finite part of the
quadratic action coming from the counterterms (except the
cosmological term already taken into account) as $S^{fin,d}_{c.t.}$.
For $d=4$ this finite piece comes only from the Einstein-Hilbert
action for $\gamma$ and, using \eqref{eq:43}, may be written
$S^{fin,d=4}_{c.t.} = - \frac{1}{16\pi G}
h_{(2)ij}G^{ijkl}h_{(2)kl}$.

The renormalized action for general even $d$ is
\begin{eqnarray}
S^{quad}_{Neu} &\approx & \frac{1}{16\pi G} \int_{\partial \cM} d^d x \big( \frac 1 4 \sum_{n=0}^{d-2} (d-n) h_{(n)ij} G^{ijkl} h_{(d-n)kl} \nonumber\\
&& +\half  \tilde h_{(d)ij} \eta^{ik}\eta^{jl} h_{(0)kl} + \alpha A^{quad}_{(d)} \big)+S^{fin,d}_{c.t.}.\label{eq:Slineven}
\end{eqnarray}
Here we have generalized the Neumann action (\ref{actionS}) to
include the additional boundary term $\frac{\alpha}{16\pi G} \int
d^d x A_{(d)} $ which
would naturally arise by changing the scale in the $\log \epsilon$
counterterm via $\log \epsilon \rightarrow \log \epsilon - \alpha$.
$A^{quad}_{(d)}$ denotes the quadratic part of $A_{(d)}$ which is given by $A^{quad}_{(4)} = -\half h_{(2)ij}G^{ijkl}h_{(2)kl}$ for $d=4$. We
have not computed this contribution in higher dimensions, though we
expect a non-zero contribution from the anomaly proportional to $C
\square^{d-4} C$ where $C$ is the Weyl tensor (see
\cite{FeffermanGraham,Parker:1987,Deser:1993yx,Karakhanyan:1996,Boulanger:2004zf,Boulanger:2007ab}).

Since $h_{ij} = O(T^{(S)}_{ij})$, the complete action (\ref{boundaryL}) that comprises the source term but no extra boundary action ($L_{diff \ inv} =0$) is finally written as
\begin{eqnarray}
S_{ren}^{(S),quad} &\approx & S^{quad}_{Neu} - \frac{1}{2}
\int_{\partial \cM}d^d x (T^{(S)}_{ij}\,h_{(0)}^{ij}-\half
\eta^{ij}T^{(S)}_{ij}\eta^{kl}\,h_{(0)}^{kl}),\label{eq:SlinS}
\end{eqnarray}
where $S^{quad}_{Neu}$ is given either in \eqref{eq:Slin} or \eqref{eq:Slineven}.

\section{Solutions of linearized gravity}

\label{lingrav}

This appendix studies linearized solutions about AdS${}_{d+1}$ in
Poincar\'e coordinates (\ref{eq:Poincare}).  The equations of motion
follow from \eqref{L2},
\begin{eqnarray}
\bar D^\alpha \bar D_\alpha h_{\mu\nu}+\bar D_\mu \bar D_\nu h -
\bar D^\alpha \bar D_\nu h_{\mu\alpha}- \bar D^\alpha \bar D_\mu
h_{\nu\alpha} - \frac{2d}{l^2}h_{\mu\nu}=0. \label{eq:linL2}
\end{eqnarray}
As in appendix \ref{app:action}, we require perturbations to satisfy
$h_{0\mu} \equiv 0$. The $0\mu$ components of \eqref{eq:linL2} can
then be integrated to yield $\eta^{ij}h_{ij} = \mathbf h = A(x^k) +
B(x^k) x_0^{-2}$, $\eta^{jk} \d_k h_{ij}-\d_i A  = C_i(x^k)
x_0^{-2}$ where $A,\,B,\,C_i$ are arbitrary functions of $x^k$.

For diffeomorphism invariant boundary theories we may further use
the boundary diffeomorphisms $\phi^i(x^k)$ to impose (see also
\cite{Waldbook}, p186) transverse gauge for the leading components
of $h_{ij} = x_0^{-2} h_{(0)ij}+O(x_0^0)$. This sets $C_i = 0$.

 The transverse mode $h_{(0)ij}$ can be decomposed into a
transverse-trace $h^{Tr,T}_{(0)ij}$ and a traceless part
$h^{TT}_{(0)ij}$. Using a boundary diffeomorphism,
$h^{Tr,T}_{(0)ij}$ is gauge equivalent to a Weyl transformation of
the boundary metric. When the boundary theory is Weyl invariant, we
may gauge fix $h^{Tr,T}_{(0)ij} = 0$. When there is no
Weyl-invariance, it is nevertheless easier to consider the (non-transverse)
perturbation induced by a Weyl transformation as this perturbation
admits the simple Fefferman-Graham expansion $h^{Weyl}_{ij}=2
\eta_{ij}\delta \sigma x_0^{-2} + \d_i \d_j \delta \sigma$. This is
done in the main text.  In the rest of this appendix we therefore
set $h^{Tr,T}_{(0)ij} = 0$. The equations of motion then imply that
$h_{ij}$ is transverse-traceless and we need only solve
\begin{equation}
\d^2_0 h_{ij}+\eta^{kl}\d_k \d_l h_{ij}+\frac{5-d}{x_0}\d_0 h_{ij} -
\frac{2(d-2)}{x_0^2}h_{ij} = 0.\label{eq:hij}
\end{equation}

We now perform a Fourier transform, $h_{ij}(x_0,x_k) = \int
\frac{d^d k}{(2\pi)^k} e^{i \eta^{ij} k_j x_i} \hat
h_{ij}(x_0,k^i)$. In the Euclidean case, $k^2 \geq 0$, there are two
independent solutions. The first one is always divergent when $x_0
\rightarrow \infty$ and must be discarded. The second solution
behaves as $x_0^{-2}+O(x_0^0)$ when $x_0 \rightarrow 0$ and may be
written in terms of the induced metric $\hat h_{(0)ij}(k^l)$  at the
boundary:
\begin{eqnarray}
\hat h_{ij}(x_0,k^l)& =& - \frac{2^{-d/2}\pi k^{d/2}}{\Gamma(d/2)e^{\frac{i d \pi}{4}} }x_0^{-2+\frac d 2} \left( Y_{d/2}[-i k x_0] +i J_{d/2}[-i k x_0]\right)\hat h_{(0)ij}(k^l)\nonumber \\
&=& \Psi^d_{Eucl}(x_0,k) \hat h_{(0)ij}(k^l),
\end{eqnarray}
where $J_n$ and $Y_n$ are the ordinary Bessel function of the first
and second kind and $k \equiv \sqrt{\eta^{ij}k_i k_j}$. When $d$ is even, the solution only depends on $k$ through $k^2$ and is regular at $x_0 \rightarrow \infty$. However, when $d$ is odd, this solution is regular at $x_0 \rightarrow \infty$ only when the positive square root $k \geq 0$ is chosen. One can read off the Fefferman-Graham coefficients from
\begin{eqnarray}
\Psi^{d=2\tilde d}_{Eucl}& =& \frac{k^{\tilde d}}{2^{\tilde d-1}\Gamma(\tilde d)} x_0^{\tilde d -2}K_{\tilde d}[k x_0]\nonumber  \\
& =& x_0^{-2} +\dots  +\frac{(-1)^{\tilde d} k^d}{2^{2\tilde d}\Gamma( \tilde d) \Gamma (\tilde d+1)}\left( H_{\tilde d} - 2(\gamma + \log(\frac{k}{2}))-\log x_0^2 \right) \nonumber\\
&& \qquad \times x_0^{d-2}  +O(x_0^{d}),\label{eq:evenpsiE}\\
\Psi^{d=1+2\tilde d}_{Eucl}&=& x_0^{-2}+\dots +\frac{(-1)^{\tilde
d+1} \pi
k^d}{2^d\Gamma{(d/2)}\Gamma{(d/2+1)}}x_0^{-2+d}+O(x_0^{-1+d}),
\label{eq:oddPsi}
\end{eqnarray}
where $K_n$ is the modified Bessel function of the second kind,
$\gamma$ is Euler's constant and $H_n = \sum_{i=1}^n \frac{1}{i}$ is
the truncated harmonic function. Expression (\ref{eq:evenpsiE}) has
been inferred by expanding $\Psi^d_{(0)}$ for each $d = 2\tilde d$
up to $\tilde d = 10$ in Mathematica, and by identifying the
harmonic function $H_{\tilde d}$ from the sequence of numbers
appearing in the expression using an engine for recognition of
integer sequences \cite{EncyclopediaofIntegerSequences}. The results
\eqref{eq:evenpsiE}-\eqref{eq:oddPsi} has been checked up to $d=20$.
For odd $d$, one finds that the (linearized) stress-tensor satisfies
\begin{equation}
\hat T_{ij} \approx  \frac{(-1)^{\frac{d+1}{2}} k^d}{2^{d+3}(\Gamma(\frac{d}{2}))^2 G} \hat h_{(0)ij}.\label{Tijodd}
\end{equation}
For $d=4$, the linearized stress-tensor \eqref{eq:Tij4} is as
\begin{equation}
\hat T_{ij} \approx  -\frac{\gamma +\frac{\alpha}{2}+ \log{(\frac k
2)}}{64 \pi G} k^4 \hat h_{(0)ij}.\label{TijEucl4}
\end{equation}

For the appropriate sources, the above $k^2 > 0$ solutions also
describe tachyons in Lorentz signature.  While (\ref{eq:oddPsi}) has
$T_{ij} \neq 0$, (\ref{TijEucl4}) vanished at $k = 2
e^{-\gamma-\frac{\alpha}{2}}$.  Thus the pure Neumann theory has no
tachyon for $d$ odd,  but does have a tachyon for $d=4$.

To find Lorentz signature propagating modes, we now consider $m^2 =
-k^2 \geq 0$. One finds two independent regular solutions.  It is
convenient to normalize these solutions at the boundary and write
\begin{eqnarray}
\Psi^d_{(d)}(x_0,m) &=& \frac{2^{d/2} \Gamma(\frac d 2 +1)}{m^{d/2}} x_0^{-2+\frac d 2} J_{d/2}[m x_0] = x_0^{d-2}+O(x_0^d),\label{defpsid}\\
\Psi^d_{(0)}(x_0,m) &=& -\frac{\pi m^{d/2}}{2^{d/2} \Gamma(\frac d 2
)} x_0^{-2+\frac d 2} Y_{d/2}[m x_0] = x_0^{-2}+O(x_0^0).
\label{defpsi0}
\end{eqnarray}
The first solution $\Psi^d_{(d)}$ is the standard normalizable mode,
while the second solution $\Psi^d_{(0)}$ changes the metric at the
boundary and is non-normalizable with respect to the standard
symplectic structure $\omega_{EH}$.

We may therefore write the general solution of \eqref{eq:hij} with
wave numbers $k^l$ as
\begin{eqnarray}
\hat h_{ij}(x_0,k^l)& =& \Psi^d_{(0)}(x_0,m) \hat
h_{(0)ij}(k^l) + \Psi^d_{(d)}(x_0,m) \hat
h^{part}_{(d)ij}(k^l),\label{generallinsol}
\end{eqnarray}
where $\hat h_{(0)ij}(k^l) $ and $\hat h^{part}_{(d)ij}(k^l)$ are
arbitrary form factors. For odd $d$, $\Psi^d_{(d)}$ is odd in $x_0$
while $\Psi^d_{(0)}$ is even. Thus $\hat h^{part}_{(d)ij}(k^l)$ is
just the Fefferman-Graham coefficient $\hat h_{(d)ij}(k^l)$.

For $d$ even, both $\Psi^d_{(d)}$ and $\Psi^d_{(0)}$ contain a term
proportional to $x_0^{d-2}$. The Fefferman-Graham coefficients $\hat
h_{(d)ij}$ (equal to $\hat h^{part}_{(d)ij}(k^l)+\cdots $) and
$\tilde {\hat {h}}_{(d)ij}$ can be deduced from the expansion
\begin{eqnarray}
\Psi^d_{(0)}(x_0,m)\hspace{-3pt} &=&\hspace{-3pt}
x_0^{-2}+\dots + \frac{H_{d/2}- 2(\gamma-\log 2 +\log{m
x_0})}{2^d \Gamma(\frac d 2)\Gamma(\frac d 2 +1)} m^d
x_0^{d-2}+O(x^d),\label{eq:expanpsi0}
\end{eqnarray}
where $H_n = \sum_{i=1}^n \frac{1}{i}$ is the harmonic function.
Again, the latter expression has been inferred from explicit
expansions (checked here up to $d=30$).

 We are now ready to summarize the solutions for Dirichlet and
Neumann boundary conditions. In any $d$, the most general solution
of the Dirichlet problem with $h_{(0)ij}=0$ is
\begin{eqnarray}
h^{Dir}_{ij}(x^0,t,\vec x) &=& \int \frac{d^d k}{(2\pi)^d}e^{-i k^1
t + i \vec k \cdot \vec x} \Psi_{(d)}^d(x^0,m)\hat
h_{(d)ij}(k^t,\vec k)\label{Dirmodes}
\end{eqnarray}
where $h_{(d)ij}(k^t,\vec k)$ is an arbitrary (transverse-traceless)
form factor. In contrast, Neumann boundary conditions require the
(linearized) stress-tensor to vanish.  Since for odd $d$ $T_{ij} =
\frac{d}{16\pi G}h_{(d)ij}$ and for $d=4$ it is given by
\eqref{eq:Tij4}, the general solution has the form
\begin{eqnarray}
h^{Neu}_{ij}(x^0,t,\vec x) &=& \int \frac{d^d k}{(2\pi)^d}e^{-i k^1
t + i \vec k \cdot \vec x} \Psi^d_{(Neu)}(x^0,m)  \hat
h_{(0)ij}(k^t,\vec k)\label{Neumodes}
\end{eqnarray}
where $ \hat h_{(0)ij}(k^t,\vec k)$ is an arbitrary
(transverse-traceless) form factor at the boundary and
\begin{eqnarray}
\Psi^d_{(Neu)}(x_0,m) & =& \Psi^d_{(0)}(x_0,m)+\alpha^{(d)}(m) \Psi^d_{(d)}(x_0,m). \label{eq:PsiT}
\end{eqnarray}
In \eqref{eq:PsiT}, $\alpha^{(d)}(m) = 0$ for odd $d$ and
\begin{equation}
\alpha^{(4)}(m) = \frac{m^4}{16}(\gamma+\log\frac{m}{2}+\frac{\alpha} 2).
\label{def:alpha4}
\end{equation}
The generalization of this solution to even $d > 4$ is
straightforward.

\section{Ghosts and positivity}
\label{sec:ghosts}

This appendix computes norm of various positive frequency modes to
determine if they are ghosts.  While the same information can be
obtained by analyzing the Euclidean propagator, we find it easier to
keep track of conventions in the Lorentz-signature calculation
below.  After fixing our conventions by checking that no ghosts are
present with Dirichlet boundary conditions, we show that timelike
ghosts never arise for Neumann boundary conditions with any $d$.
However, the tachyonic solutions for even $d \ge 4$ will lead to
ghosts.

The key point is that the symplectic form \eqref{eq:defomegaform}
defines a natural inner product
\begin{equation}
\langle \delta_1 g, \delta_2 g \rangle \equiv i \int_\Sigma
\omega_{Neu}[\delta_2 g^*,\delta_1 g].
\end{equation}
Around the flat background \eqref{eq:Poincare} and for perturbations
in Fefferman-Graham and transverse-traceless gauge, the symplectic
structure~\eqref{omegaEHgamma} on $x^1 \equiv t$ constant slices
reduces to,
\begin{equation}
\omega^t_{EH}[\delta_2 g^{TT}, \delta_1 g^{TT}] =
\frac{x_0^{-d+5}}{32\pi G}(\delta_2
g^{TT}_{ij}\eta^{ik}\eta^{il}\d_t \delta_1 g^{TT}_{kl} -
(1\leftrightarrow 2)),\label{symt}
\end{equation}
where we have neglected boundary terms which will vanish for wave
packets $\delta g_{kl}$ decreasing sufficiently fast at Poincar\'e
infinity $\vec x \equiv x^2,\dots,x^{d-1}\rightarrow \infty$.

Let us first consider the norm of a solution with Dirichlet boundary
conditions~\eqref{Dirmodes}. Since the counterterms make no
contribution to the symplectic structure, we have $\omega_{Neu} =
\omega_{EH}$.  In that case, the inner product reduces to the
closure relation for Bessel functions of the first kind:
\begin{eqnarray}
\int_0^\infty \frac{dx^0}{x_0^{d-5}} \Psi_{(d)}^d(x_0,m)
\Psi_{(d)}^d(x^0,m^\prime) &=&
\frac{2^d (\Gamma(\frac d 2 +1))^2}{m^{\frac d 2}m^{\prime \frac d 2}}\int_0^\infty dx^0 x^0 J_{\frac d 2}(m x^0)  J_{\frac d 2}(m^\prime x^0)\nonumber\\
&=&\kappa^2_{(d)}(m)  \delta(m -
m^\prime),\label{closureJ}
\end{eqnarray}
where $\kappa^2_{(d)}(m) = \frac{2^d (\Gamma(\frac d 2
+1))^2}{m^{d+1}}$.  Here we have used $m
>0$, $m^\prime >0$ \cite{Arfkenbook}. Using the sign convention
$\int_\Sigma \omega_{Neu} \equiv \int^\infty_0 dx^0 \int d\vec x
\omega^t_{Neu}$, the inner product becomes
\begin{equation}
\langle h^{Dir}, h^{Dir} \rangle = \frac{1}{16\pi G}  \int
\frac{m \kappa^2_{(d)} d^d k }{(2\pi)^d} \text{sign}(k^1)
h^*_{(d)ij}(k^l)  h_{(d)}^{\;\;ij}(k^l),
\end{equation}
where we have used \eqref{closureJ} and we assume that the form
factors vanish sufficiently fast as one approaches $m
\rightarrow 0$ in order to have a finite integral. This expression
is manifestly positive for positive frequency modes since
\begin{equation}
h^*_{(d)ij}(k^l)  h_{(d)}^{\;\;ij}(k^l) \geq 0 \label{positivef}
\end{equation}
for transverse tensors with $k^i$ timelike.

Let us now turn to the inner product between two Neumann
modes~\eqref{Neumodes}. Since the symplectic structure
$\omega_{Neu}$ is conserved, modes with different $m$ must be
orthogonal.  (If not, the norm of a superposition of modes would not
be conserved.) We must merely identify the analogue of the closure
relation \eqref{closureJ} for the Bessel functions \eqref{eq:PsiT}.
It is clear that contributions from the counterterms is crucial. For
odd $d = 2 \tilde d+1$, there are $\tilde d + 1$ counterterms
(though the cosmological constant counterterm does not affect the
symplectic structure). These must cancel $\tilde d$ divergent terms
in $\omega_{EH}$.  For even $d=2\tilde d$, there are $\tilde d-1$
polynomial divergences and one logarithmic divergence in the
(integrated) symplectic structure. In summary, there are always
$\tilde d$ polynomial divergences in the symplectic form, one which
is proportional to $1/x_0$ in the even $d$ case.

It is straightforward to see that exactly the same patterns occurs
in the integrand
\begin{eqnarray}
(x^0)^{-d+5} \Psi_{(Neu)}^d(x_0,m) \Psi_{(Neu)}^d(x^0,m^\prime) =
c^{d}_{1}(m,m^\prime) x_0^{-d+1} +
c^{d}_{2}(m,m^\prime) x_0^{-d+3} + \dots  \label{eq:divpattern}
\end{eqnarray}
By our choice of normalization, $c^{d}_{1}(m,m^\prime) = 1$.
Evaluating the left-hand side explicitly, we obtain the next
coefficients:
\begin{eqnarray}
c^{d}_{2}(m,m^\prime) = \frac{m^2+m^{\prime
2}}{2(d-2)}, \qquad c^{d}_{3}(m,m^\prime) =
\frac{m^4+m^{\prime 4}}{8(d-2)(d-4)}+\frac{m^2
m^{\prime 2}}{4(d-2)^2},\quad  \dots \label{coefc}
\end{eqnarray}

Let us discuss first the odd $d$ case. The functions
$\Psi_{(Neu)}^d(x_0,m) = \Psi_{(0)}^d(x_0,m)$ involve only Bessel
functions of the second kind and we expect the closure relation to
be
\begin{eqnarray}
&&\int_0^\infty dx^0 \left[ (x^0)^{-d+5} \Psi_{(0)}^d(x_0,m) \Psi_{(0)}^d(x^0,m^\prime) - \sum_{n=1}^{\tilde d} c^{d}_{n}(m,m^\prime) x_0^{-d+1+2 n} \right] \nonumber \\
&&=\kappa^2_{(0)}(m) \delta(m - m^\prime)\label{closuredodd}
\end{eqnarray}
where the coefficients $\kappa^2_{(0)}$ should be computed. By
imposing a cut-off $N$ on the upper limit of integration, and using
the representation of the delta pseudo-function $\delta(m -
m^\prime) = \text{lim}_{N\rightarrow \infty} \frac{\sin((m -
m^\prime)N)}{\pi (m - m^\prime)}$, we have used Mathematica to check
symbolically that this formula is valid for odd $d \le 9$ with
\begin{eqnarray}
\kappa^2_{(0)} (m)= \frac{\pi^2 }{2^d (\Gamma(\frac{d}{2}))^2
\,}m^{d-1}.\label{kappa}
\end{eqnarray}

In even dimensions, the closure relation is more involved. One
complication is that  the modes $\Psi_{(0)}^d(x_0,m)$ and
$\Psi_{(d)}^d(x_0,m)$ are not orthogonal,
\begin{eqnarray}
\int_0^\infty dx^0  (x^0)^{-d+5} \Psi_{(d)}^d(x_0,m)
\Psi_{(0)}^d(x^0,m^\prime) =\frac{d}{m^2-m^{\prime
2}}.\label{cross}
\end{eqnarray}
This cross-term will be weighted by $\alpha^{(d)}(m)-\alpha^{(d)}(m^\prime)$, see \eqref{eq:PsiT}.

The inner product of two modes $\Psi_{(d)}^d(x_0,m)$ is given by
\eqref{closureJ}.  It remains to calculate the inner product of two
modes $\Psi_{(0)}^d(x_0,m)$.  As already pointed out in
\eqref{eq:43}, for even $d = 2 \tilde d$ the counterterms contribute
a finite piece to the symplectic structure in addition to canceling
(both power law and logarithmic) divergences. Thus the closure
relation must take the form
\begin{eqnarray}
&&\int_{\sqrt{\eps}}^\infty dx^0 \left[ (x^0)^{-d+5} \Psi_{(Neu)}^d(x_0,m) \Psi_{(Neu)}^d(x^0,m^\prime) - \sum_{n=1}^{\tilde d - 1} c^{d}_{n}(m,m^\prime) x_0^{1-2 \tilde d+2 n} \right] +\half \log{\eps} \,c^{d}_{\tilde d} \nonumber \\
&&-F_{c.t.}^{(d)}(m,m^\prime)
=(\kappa^2_{(0)}+\alpha_{(d)}^2(m) \kappa^2_{(d)})
\delta(m - m^\prime),\label{closuredeven}
\end{eqnarray}
where the equality holds in the sense of distributions; i.e.,  when
it is multiplied by a function that decreases sufficiently fast as
$m\rightarrow 0$ (in our regularization of the integral, we
encountered expressions of the type $\text {lim}_{N \rightarrow
\infty} \frac{\cos(N\,m)}{m}$).

The coefficients in front of the divergent terms are obtained from
the expansion \eqref{eq:divpattern} and the requirement that
divergences cancel. Here $F_{c.t.}^{(d)}(m,m^\prime)$ is a
polynomial in $m$ and $m^\prime$ which represents the finite
contribution from the  counterterms.  This function is zero for $d$
odd and $d=2$. For $d=4$ the only contribution comes from the
Einstein-Hilbert term, whose contribution to the symplectic
structure can be obtained from the relation between the symplectic
structures  \eqref{eq:omegaEHgamma}-\eqref{symt}-\eqref{omega4} and
from the Fefferman-Graham expansion of the solution \eqref{defpsi0}
(in particular $h_{(2)ij}(m) = \frac{m^2}{4} h_{(0)ij}(m)$. One
finds $F_{c.t.}^{(4)} = \frac{m^2+m^{\prime 2}}{8}$. This allows us
to check that all finite terms (that are not delta functions)
cancel, as must occur for the symplectic structure to be conserved.
The coefficients $\kappa^2_{(0)}$ and $\kappa^2_{(d)}$ are given
below \eqref{closureJ} and in \eqref{kappa}. They are obtained by
the regularization procedure explained above. The coefficients
$c_{d/2}^d$ can be deduced from the expansion \eqref{coefc} and are
given by
\begin{equation}
c_{d/2}^d = \frac{4}{(\Gamma(\frac d 2))^2 2^d}\frac{m^d -
m^{\prime d}}{m^2 - m^{\prime 2}}.
\end{equation}

The norm of a general Neumann perturbation follows quickly from
(\ref{closuredeven}).  Using the analogue of \eqref{positivef} for $h_{(0)ij}$, we see that the
result
\begin{equation}
\langle h^{Neu}, h^{Neu} \rangle =  \int \frac{d^d k}{(2\pi)^d}
\frac{(\kappa^2_{(0)}+\alpha_{(d)}^2(m) \kappa^2_{(d)})}{16\pi G}
\text{sign}(k^1) h^*_{(0)ij}(k^l)  h_{(0)}^{\;\;ij}(k^l)
\end{equation}
is non-negative for positive frequency modes when $k^i$ is timelike.
It follows that the Neumann theory has no ghosts with $\sqrt{-k^2} =
m^2 \geq 0$. This completes the analysis for odd $d$.  On the other
hand, Neumann tachyons arise for even $d \ge 4$.  For tachyons,
$k^i$ is spacelike and the sign of $h^*_{(0)ij}(k^l)
h_{(0)}^{\;\;ij}(k^l)$ depends on the polarization of the mode; both
ghost and normal polarizations occur.

Note that the form of \eqref{closuredeven} provides a non-trivial
check on our calculations. The change $\log\eps \rightarrow \log\eps
- \alpha$ adds a finite term to the Lagrangian which changes the
definition of the stress-tensor, affecting the value of
$\alpha^{(d)}(m)$. But the logarithmic term in the symplectic
structure, and therefore in \eqref{closuredeven}, also shifts. Since
the symplectic structure remains conserved, modes with different
values of $m$ must remain orthogonal and the cross-term coming from
the integrals \eqref{cross} must cancel against the shift of the
counterterm. This requires
\begin{equation}
\alpha^{(d)}(m)|_{\alpha} = \alpha^{(d)}(m)|_{\alpha=0} +
\frac{m^d}{d (\Gamma(\frac d 2))^2 2^{d-1}}\alpha,
\end{equation}
which provides a non-trivial check of the $\alpha$ dependence in $\eqref{def:alpha4}$ for $d=4$.


\providecommand{\href}[2]{#2}\begingroup\raggedright\endgroup

\end{document}